\documentclass[prd,twocolumn,showpacs,nofootinbib,amsmath,amssymb,floatfix,superscriptaddress,showkeys]{revtex4}
\usepackage{multirow}
\usepackage{graphicx}
\usepackage{epstopdf}
\usepackage{dcolumn}
\usepackage{bm}
\usepackage{threeparttable}
\usepackage{subfigure}
\usepackage{color,txfonts}
\usepackage{ulem}
\definecolor{blue0}{rgb}{0,0,0.6}
\usepackage[colorlinks,linkcolor=blue0,anchorcolor=blue0,citecolor=blue0,urlcolor=blue0]{hyperref}

\newcommand{\jcap}{J. Cosmol. Astropart. Phys.}
\newcommand{\physrep}{Phys. Rep.}
\newcommand{\mnras}{Mon. Not. R. Astron. Soc.}
\newcommand{\araa}{Annu. Rev. Astron. Astrophys.}
\newcommand{\aap}{Astron. Astrophys}

\newcommand{\beq}{\begin{equation}}
\newcommand{\eeq}{\end{equation}}
\newcommand{\beqa}{\begin{eqnarray}}
\newcommand{\eeqa}{\end{eqnarray}}

\begin{document}

\title{The prediction of using LHAASO's cosmic-ray electron measurements to constrain decaying heavy dark matter}

\author{Ben-Yang Zhu}
\affiliation{Laboratory for Relativistic Astrophysics, Department of Physics, Guangxi University, Nanning 530004, China}
\author{Yun-Feng Liang}
\email[]{liangyf@gxu.edu.cn}
\affiliation{Laboratory for Relativistic Astrophysics, Department of Physics, Guangxi University, Nanning 530004, China}

\date{\today}

\begin{abstract}
LHAASO is an instrument designed for detecting cosmic rays (CRs) and gamma rays at TeV to PeV energies. The decays of heavy dark matter particles in the Galactic halo may produce high-energy electrons that can be detected by LHAASO. The main background for the LHAASO's CR electron {measurements is the} hadron residuals due to mis-identification of the particle species.
In this paper, we estimate the LHAASO's electron background using the known all-particle CR spectrum and the hadron rejection efficiency of LHAASO. With the estimated background, we predict the capability of LHAASO to constrain DM decay lifetime at $95\%$ confidence level for various channels.
We find that, if neglecting systematic uncertainties, the CR electron measurement by LHAASO can improve the current best results by up to one order of magnitude for DM masses between $100-1000\,{\rm TeV}$. However, indirect measurements of CR electrons by ground-based experiments suffer from sizeable systematic uncertainties. With the systematic uncertainties included in the calculation, the projected constraints will be largely weakened.
So for using the CR electron observation of LHAASO to constrain the DM parameters, the key point is whether the systematic error can be effectively reduced.
\end{abstract}

\maketitle

\section{Introduction}

Baryonic matter only accounts for $\sim5\%$ energy density of the known universe \cite{Planck2020}. Most of the matter in the universe exists in the form of the so-called dark matter (DM), of which the nature is still unknown. The DM particles account for about 27\% of the total energy density of the Universe \cite{Planck2020}. Their existence is supported by many astrophysical and cosmological phenomena such as galaxies' rotation curves and gravitational lensing of galaxy clusters \cite{1996PhR...267..195J,2005PhR...405..279B}. In general, we consider the dark matter to be particles beyond the Standard Model (SM) {(note that there exists another class of theories to explain the dark matter problem without the need of new particles called MOdified Newtonian Dynamics or MOND \cite{mondrv02,mondcmb21})}. The search for such particles is an important topic in today’s physics and astronomy research.

Weakly interacting massive particle (WIMP) is one of the most popular dark matter candidates. This class of particles may interact with SM particles in the energy scale of weak interaction. They may also self-annihilation or decay into SM particles, producing detectable signals such as $\gamma$-rays, electrons/positrons, neutrinos \cite{1996PhR...267..195J,2005PhR...405..279B}. So we can search for the existence of DM or constrain its parameters through these indirect signals (the so-called indirect DM detection). However, although many efforts have been paid to search for DM signals based on the observations from a variety of instruments (such as AMS-02 \cite{ams13positron,2014PhLB..728..250F,2014PhRvD..89f3539I,2017PhRvL.118s1101C,2017PhRvL.118s1102C}, DAMPE \cite{2017APh....95....6C,DAMPE:2017fbg,Yuan:2017ysv,fan18dampe,dampe22line,ltc22dampe}, Fermi-LAT \cite{Atwood2009,hooper11gce,weniger12line,bringmann12line,gordon13gce,fermi15line,fermi15dsph,zhou15gce,Charles16,fermi17dsph,fermi17gce,ls21dsph}, HAWC \cite{HAWC:2014ycj,HAWC:2017mfa,HAWC:2017udy}), no robust DM signal is detected so far.

Decaying heavy dark matter has been proposed in many literature \cite{lhaaso22decay}. One example is the decaying gravitino in supergravity model \cite{Ibarra:2007wg,Ishiwata:2008cu,Cohen:2016uyg,Dudas:2018npp}. Other interesting candidates include WIMPzillas \cite{Kolb:1998ki} and glueballs \cite{Cohen:2016uyg,Halverson:2016nfq}. Based on gamma-ray and neutrino data, heavy dark matter has been extensively studied in a very wide mass range \cite{Cohen:2016uyg,Kalashev:2016cre,IceCube:2018tkk,Kachelriess:2018rty,Bhattacharya:2019ucd,Ishiwata:2019aet,Chianese:2021jke,IceCube:2022vtr,lhaaso22decay}. The heavy DM has also been proposed to explain the diffuse TeV-PeV neutrino spectrum observed by IceCube \cite{Cohen:2016uyg,Chianese:2016kpu,Kachelriess:2018rty}.

In recent years, many new instruments for the observation of very-high-energy $\gamma$-rays and cosmic rays have been constructed / are under construction, such as the new generation of ground-based Cherenkov detectors LHAASO \cite{bai2022large} and CTA \cite{CTAConsortium:2013ofs}. These detectors would improve our ability to detect TeV-PeV gamma-ray and cosmic-ray signals from cosmic space. Thanks to these instruments, we can perform more powerful searches for electrons/positrons and $\gamma$-rays produced by dark matter particles with energies over 100 TeV, which may help to study heavy dark matter particles or give stronger constraints on their parameters. For a review of the prospects of DM searches with LHAASO, one can see Ref.~\cite{lhaaso21_swpch5}.

The main particle species detected by LHAASO is VHE gamma rays. 
Some significant progress has been made in the observations of ultra-high-energy gamma rays \cite{2021Natur.594...33C,LHAASO:2021cbz}.
Based on gamma-ray data, observations of dwarf galaxies, the Galactic halo and the Galactic center can be used to search for dark matter signals. 
Using 570 days of gamma-ray observations, LHAASO has provided some strongest constraints on the lifetime of heavy dark matter particles with masses between $10^5$ and $10^9$~GeV \cite{lhaaso22decay}.
LHAASO also measures spectra of electron and proton cosmic rays (CRs) \footnote{Because the LHAASO measurement can not discriminate between particles ($e^-$, $p$) and antiparticles ($e^+$, $\bar{p}$), the terms electron/proton is used to refer to the sum of both particle and antiparticle in this paper.}. In this work, we focus on LHAASO's CR detection. 
The possibility of using HAWC or LHAASO to detect CR electrons (CREs) and constrain DM parameters has been suggested by Ref.~\cite{zhou17solar}.
They showed that HAWC and LHAASO can provide unprecedented sensitivity to the 5–70 TeV CR electron spectrum, which allows interesting tests of dark matter. 
The LHAASO detection of CR electrons is mainly limited by the hadron background (i.e. the hadron contamination of electron data due to electron/hadron separation efficiency). Based on the currently known CR hadron spectrum and the LHAASO's hadron-rejection power, we estimate the detection threshold of LHAASO for CR electrons. According to the threshold we give an estimation of the lower limits that the future LHAASO CR electron observations can place on the lifetime of decaying heavy DM.

\begin{figure}
\centering
\includegraphics[width=0.45\textwidth]{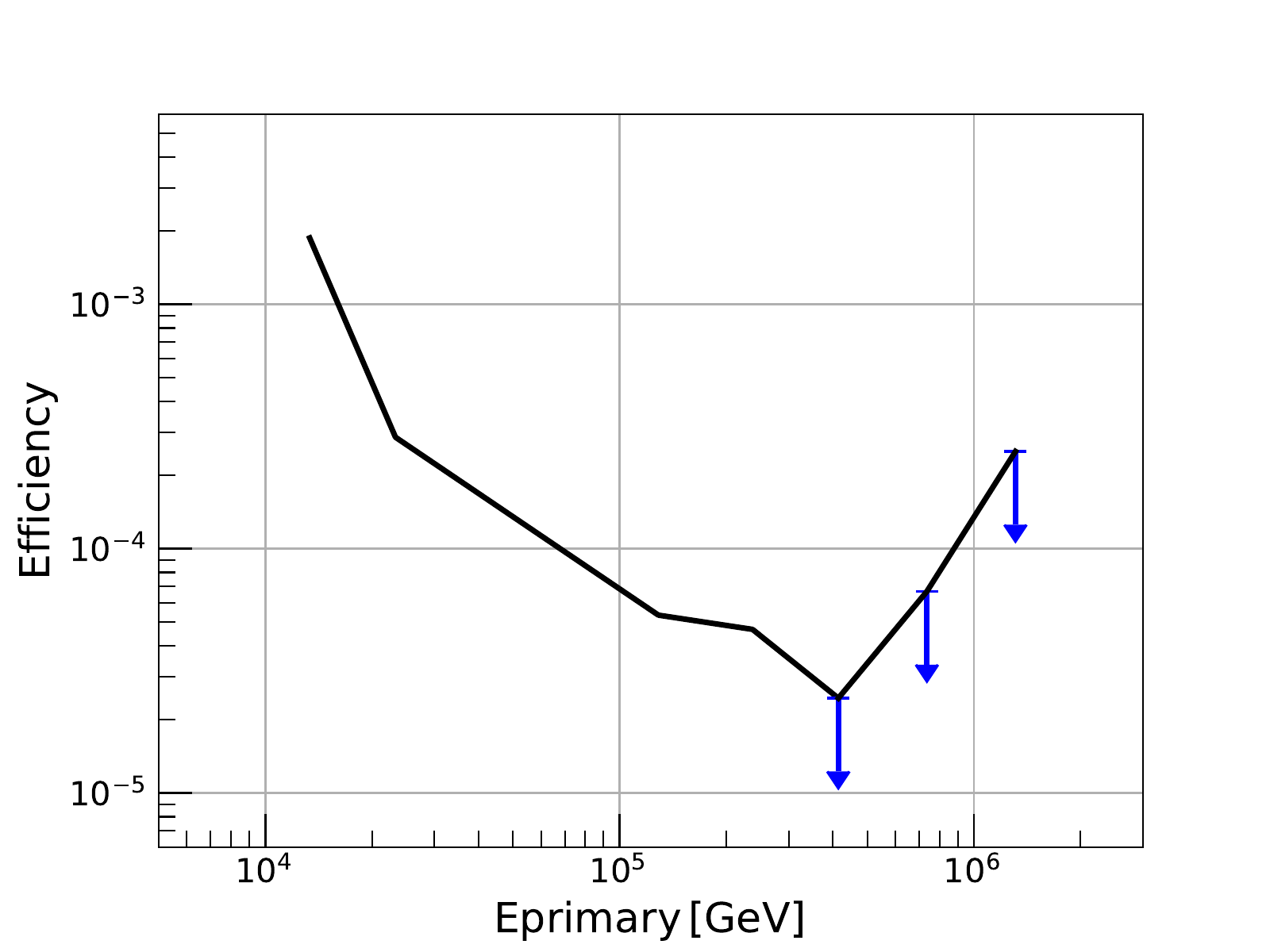}
\caption{Background rejection power of hadron CR of LHAASO \cite{bai2022large}. This curve means only $\sim 10^{-4}-10^{-3}$ hadron CR particles will be mis-identified as electrons/photons after the data selection. {In the simulation of deriving this curve, the electromagnetic particle detector (ED), muon detector (MD), and WCDA components of LHAASO are included.}}
\label{fig:rej}
\end{figure}

\section{Analysis} 

\subsection{The LHAASO experiment and lepton/hadron separation}

The Large High Altitude Air Shower Observatory (LHAASO) \cite{bai2022large} is an air shower cosmic-ray and $\gamma$-ray detector located in southwestern China. It mainly consists of KM2A (Kilometer Square Array), WCDA (Water Cherenkov Detector Array) and WFCTA (Wide Field of view Cherenkov Telescope Array). The full array of LHAASO has been completed in 2021. The LHAASO is sensitive to $\gamma$-rays above 100 GeV. It for the first time detected PeV $\gamma$-rays from astrophysical sources \cite{2021Natur.594...33C}. The LHAASO can also detect electron and hadron cosmic rays through the secondary
particles generated in extensive air shower and may extend the measurements of CR electron spectrum into PeV energies.

Most of the events recorded by LHAASO are showers induced by hadron cosmic rays, which constitute the main background for the observations of $\gamma$-rays / electron cosmic rays. Considering that $\gamma$-ray-induced showers are muon-poor and hadron CR induced is muon-rich, the hadron CR background can be suppressed very effectively by measuring the muon component in the showers \cite{lhaaso2021crabperf}. The efficiency of the hadron rejection cuts for gamma rays and protons is shown in Fig.~\ref{fig:rej}. {The fraction of hadron CRs that survive the rejection cuts is $\sim0.02\%$ at 30 TeV, while at 500 TeV, it is found to be $<$0.004\% \cite{bai2022large,lhaaso21_swpch2}.} This curve of rejection efficiency for hadron cosmic rays will be used in our subsequent analysis.
\begin{figure}[b]
\centering
\includegraphics[width=0.45\textwidth]{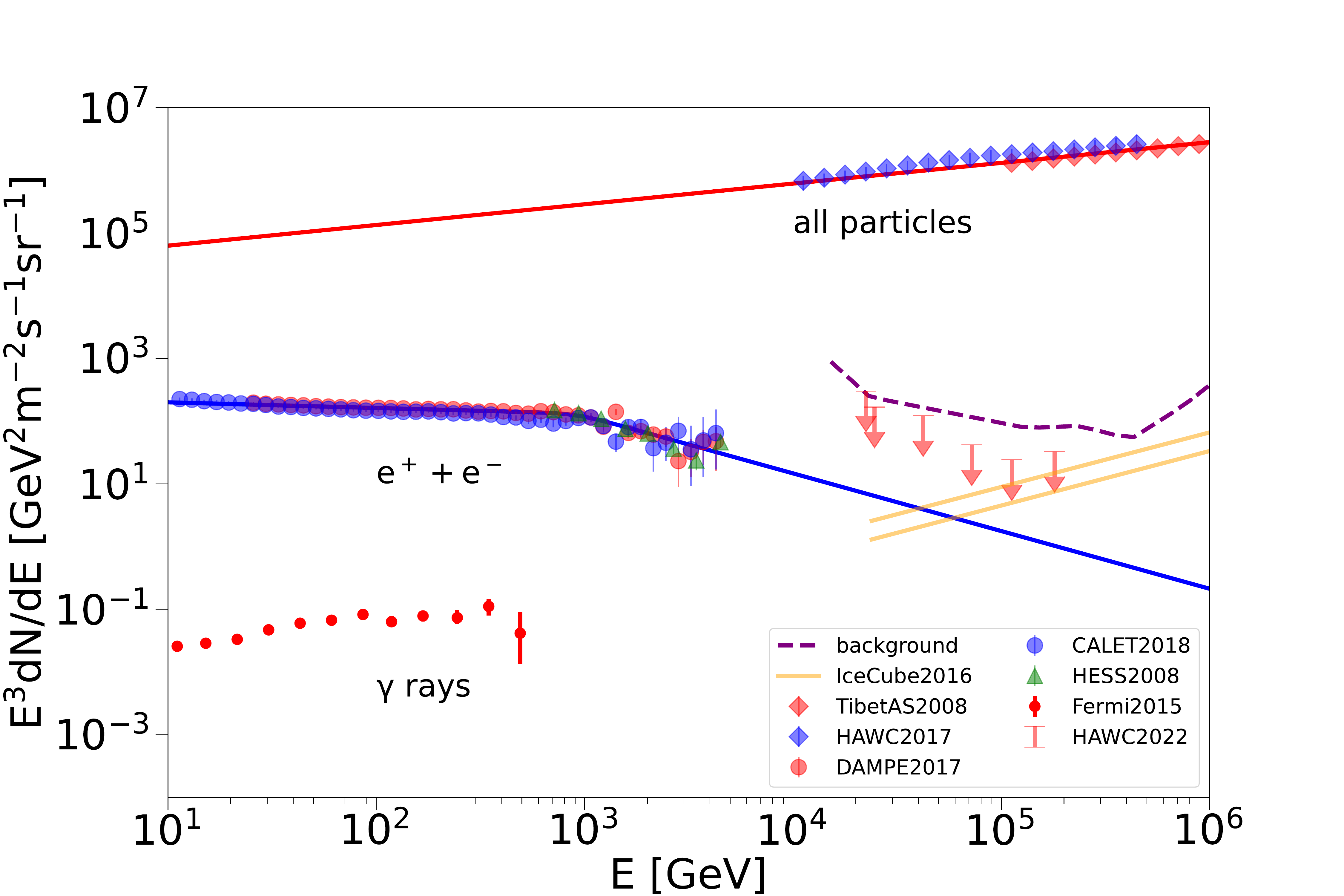}
\caption{CR spectra of electrons by DAMPE (red circles) \cite{DAMPE:2017fbg}, CALET (blue circles) \cite{Adriani:2018ktz} {and HESS (green triangles) \cite{HESS:2008ibn}}, and all-particles by {Tibet AS} (red diamonds) \cite{tibet08} and HAWC (blue diamonds) \cite{Morales-Soto:2022qzw}, together with the best-fit lines to the spectra (red and blue solid lines). The purple dashed line is the estimated background for LHAASO's CR electron measurements. Also shown is the EGRB measurements by Fermi-LAT \cite{Fermi-LAT:2014ryh} and upper limits by HAWC \cite{HAWC:2022uka}. The two orange lines are the gamma-ray flux corresponding to the IceCube diffuse neutrino flux \cite{IceCube:2016umi} for both $p\gamma$ and $pp$ interactions.}
\label{fig2}
\end{figure}

\subsection{Energy spectrum of CR electron and hadron}

The spectrum of hadron CRs can be represented by the CR all-particle spectrum. At the energies above 10 TeV, the spectrum is mainly measured through detecting secondary particles in the extensive air showers produced by the CR interaction with nuclei in the Earth’s atmosphere. The CR spectrum above 10 TeV has been measured by many instruments, such as HAWC \cite{Morales-Soto:2022qzw}, Icecube-Icetop \cite{IceCube:2019hmk}, KASCADE-Grande \cite{2012APh....36..183A} and Tibet AS \cite{tibet08}. The all-particle spectrum follows a power-law function before $\sim3\times10^3\,{\rm TeV}$ (‘knee’) with a spectral index $\alpha\sim-2.7$. The spectrum breaks at $\sim3\times10^3\,{\rm TeV}$ and the power-law index becomes steeper ($\alpha\sim-3.0$) above this energy. 

{Among all the all-particle spectra, the measurements from Tibet AS \cite{tibet08} and HAWC \cite{Morales-Soto:2022qzw} cover the most energies we are interested in. However, there is a systematic deviation in the fluxes of the two (at a level of $\sim$20\%-30\%), so we do not fit them simultaneously and adopt only the flux points of Tibet AS.}
We fit the all-particle spectrum observed by Tibet AS (QGSJET+HD in \cite{tibet08}) in the energy range from 10- 500 TeV with a single power-law function, ${\Phi = \Phi_0\times (E / 
100\,\mathrm{GeV})^{-\gamma}}$ . The best-fit parameters are $\Phi_0$ = $1.34\times 10^{-1} \mathrm{m}^{-2} \mathrm{s}^{-1} \mathrm{sr}^{-1} \mathrm{GeV}^{-1}$ and $\mathrm{\gamma} = -2.64$ , which will be used and extended to the LHAASO's whole energy range in the later analysis. We have also tested that adopting the HAWC spectrum \cite{Morales-Soto:2022qzw} would not affect the final results too much.

For electron CRs, at present the most precise measurement of the spectrum at TeV energies is from the DAMPE satellite, which gives the measurement up to the energy of 4.6 TeV \cite{DAMPE:2017fbg}. 
DAMPE has an unprecedentedly high energy resolution and low hadronic cosmic-ray background. The DAMPE electron CR spectrum shows a spectral hardening near 50 GeV and a spectral break near 0.9 TeV. Above 55 GeV, the spectrum can be well fitted by a smoothly broken power-law (SBKPL) model, $\Phi=\Phi_0(E / 100\,\mathrm{GeV})^{-\gamma_1}\left[1+\left(E / E_{\mathrm{b}}\right)^{-\left(\gamma_1-\gamma_2\right) / 0.1}\right]^{-0.1}$. The best-fit parameters are $\gamma_1=3.09 \pm 0.01$, $\gamma_2=3.92 \pm 0.20$, $\Phi_0=(1.62 \pm 0.01) \times 10^{-4}~\mathrm{m}^{-2} \mathrm{s}^{-1} \mathrm{sr}^{-1} \mathrm{GeV}^{-1}$ and $E_{\mathrm{b}}=914 \pm 98\,\mathrm{GeV}$ {(the best-fit parameters are from \cite{DAMPE:2017fbg} and result in $\chi^2/{\rm d.o.f.}=23.3/18$)}. We show the all-particle and electron spectra in Fig.~\ref{fig2}.

At higher energies, the direct measurement of electrons by space-based detectors is difficult due to the limited statistics. The ground-based experiments such as LHAASO are therefore expected to give important information on the CR electrons at high ($>10$ TeV) energies. Here we directly extrapolate the best-fit SBKPL model of DAMPE data into higher energies to give an estimation of the intrinsic spectrum of CR electrons in LHAASO's energy range.
We find that for the LHAASO observation, the electron background for DM search is not dominated by the intrinsic astrophysical electrons, since the flux of the astrophysical component is expected to be very low at these energies (see the extrapolation of the CRE spectrum in Fig.~\ref{fig2}).

\subsection{Background for DM search in electron CRs}
\label{sec:bkg}
The main background results from the mis-identified cosmic-ray particles due to the efficiency of the $e/p$ separation. Using the rejection efficiency and the all-particle spectrum we obtain the expected background for DM search in electron CRs. The background can be expressed as 
\begin{equation}
F_{\rm bkg}(E)=\Phi_p(E)\times \eta(E)
\label{eq:fbkg}
\end{equation}
where the $\eta$ and $\Phi_p(E)$ represent the hadron rejection efficiency and the CR all-particle spectrum, respectively.
{Note that we have assumed the reconstructed energy of a particle is equal to its intrinsic energy ($E= E_{\rm rec}\approx E_{\rm true}$, i.e., ignoring the energy dispersion). This may bias the final results by tens of percent.
In addition, by introducing the $E_{\rm rec}$, we are in fact using the convention that is more commonly used for gamma-ray data analysis.
For the analysis of CRs, people are often not concerned with the reconstructed energies of individual particles, but with the distribution of shower parameters $\vec{s}$. However, since it is always possible to give a particle a reconstructed energy according to the $\vec{s}$, our prescription here is also valid for LHAASO CR observations.
Another caveat is that, at the same intrinsic particle energy, the shower size of hadron and electron are different \cite{KASCADEGrande:2017vwf}. The shower of electron is larger. So at the energy of $E_{\rm rec}$ , the residual hadrons in the CR electrons have an intrinsic energy of $E_{\rm hadron,true}>E_{\rm rec}$ (note $E_{\rm electron,true}\approx E_{\rm rec}$).
The Eq.~(\ref{eq:fbkg}) should be $F_{\rm bkg}(E_{\rm obs})=\Phi_p(E_{\rm hadron,true}(E_{\rm obs}))×\eta(E_{\rm obs})$.
However, considering that $\Phi$ has a spectral index of $<0$, the approximate treatment of assuming $E_{\rm hadron,true}\approx E_{\rm rec}$ in Eq.~(\ref{eq:fbkg}) is actually conservative.}

The background spectrum based on Eq.~(\ref{eq:fbkg}) is shown in Fig.~\ref{fig2} (purple dashed line). We can see that this instrument background is higher than the intrinsic CR electrons by $\sim2$ orders of magnitude so that the latter can be safely neglected.

{The gamma-ray photons are also not expected to contribute a lot to the background. For the gamma rays from point sources and the diffuse emission from the Galactic plane, they can be subtracted from electrons using the directional information. {Subtracting the data from the Galactic plane region only reduces the event statistics of electrons by at most $\sim20\%$ \cite{LHAASO:2023gne}, which has little impact on the final results.} The only gamma-ray component that may contaminate the CR electron measurements is the ones from the unresolved extragalactic sources (i.e., the isotropic extragalactic gamma-ray background, EGRB). Currently, we do not have direct measurements for EGRB above 1 TeV \cite{Fermi-LAT:2014ryh,HAWC:2022uka}. However, some works have derived upper limits on the EGRB flux, and in Fig.~\ref{fig2} we also show the flux upper limits given by Refs.~\cite{HAWC:2022uka} and \cite{Inoue:2012cs}. We can see that the upper limits are below the expected background (dashed purple line). In addition, if the TeV EGRB emission is produced by $p\gamma$ or $pp$ interactions, the flux (upper limits) can be estimated through the observed diffuse neutrino flux of IceCube \cite{IceCube:2016umi,HAWC:2022uka}. The intrinsic/unattenuated gamma-ray flux (orange lines in Fig.~\ref{fig2}) derived from the IceCube diffuse neutrino flux is already lower than our expected background. In practice, the realistic/observed EGRB flux is likely to be much lower than this, because it is subject to absorption via pair-production on the extragalactic background light (EBL) \cite{Liang:2020roo}. All these facts suggest that the background from gamma rays is negligible for LHAASO electron measurements.} 

\begin{figure}
\centering
\includegraphics[width=0.49\textwidth]{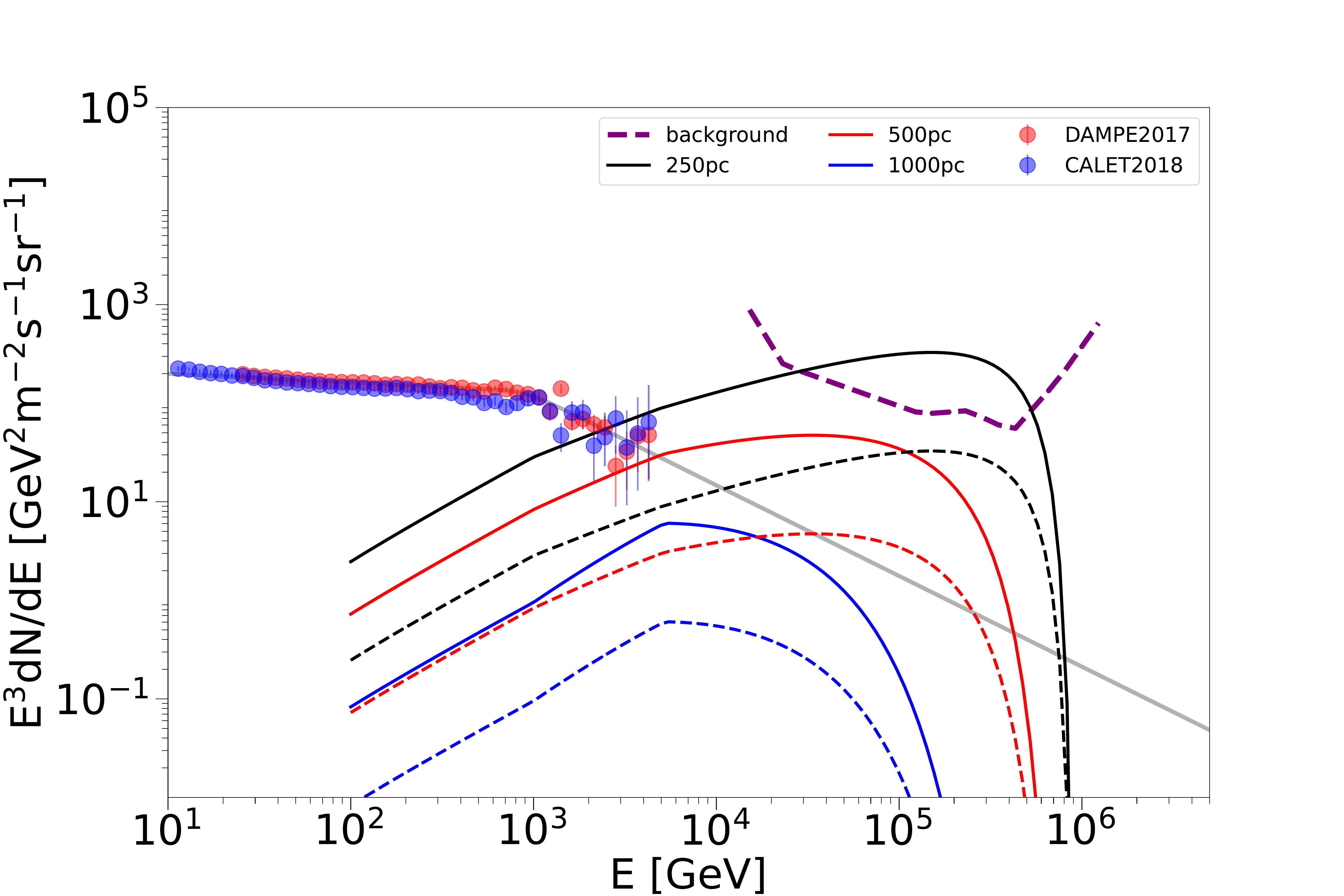}
\caption{Possible contribution from nearby astrophysical sources to the TeV-PeV electron spectrum for different injection luminosities and source distances. The solid and dashed lines are for injection luminosities of $10^{34}\,{\rm erg\,s^{-1}}$ and $10^{35}\,{\rm erg\,s^{-1}}$, respectively.}.
\label{fig:nearsrc}
\end{figure}

{Another point worth noting is that, nearby astrophysical sources may contribute to the CR electron spectrum of LHAASO. This astrophysical component may create possible bump in the electron spectrum and elevate the background for DM search. Here we estimate the electron flux in the LHAASO's energy band caused by the nearby astrophysical sources. Possible acceleration sources of Galactic TeV cosmic rays include supernova remnants (SNRs) and pulsar wind nebulae (PWNs), which have a typical injection power of $10^{33}-10^{35}\,{\rm erg/s}$ \cite{Acero:2013xta,Xin:2018vut}. The minimum distance of the currently known TeV SNRs and PWNs is $\sim$250 pc \footnote{\url{http://tevcat.uchicago.edu/}}. With this information we estimate the CR electron flux contributed by the nearby sources, as is shown in Fig.~\ref{fig:nearsrc} (the details of the calculation can be found in Appendix~\ref{app:a}).
In our calculation we have considered an injected electron spectrum with $\gamma=2.0$ and $E_{\rm cut}=1000\,{\rm TeV}$. It can be seen that if requiring the flux of the resulting electron spectrum not contradict the DAMPE and CALET observations, the contribution from these sources in the LHAASO's energy bands will not exceed our expected electron background. However, it should be noted that if there exist unknown closer (e.g., 100 pc) injection sources, they may produce CR electrons exceeding the background and are detectable by LHAASO.
Nevertheless, if we do find a component beyond the expected background, whether an astrophysical or dark matter origin, it is an important discovery. In the following, we will not consider the existence of a nearby astrophysical component and assume the LHAASO background for electrons can be well represented by the purple line of Fig.~\ref{fig2}.}

\begin{figure}
\label{fig3}
\centering
\includegraphics[width=0.45\textwidth]{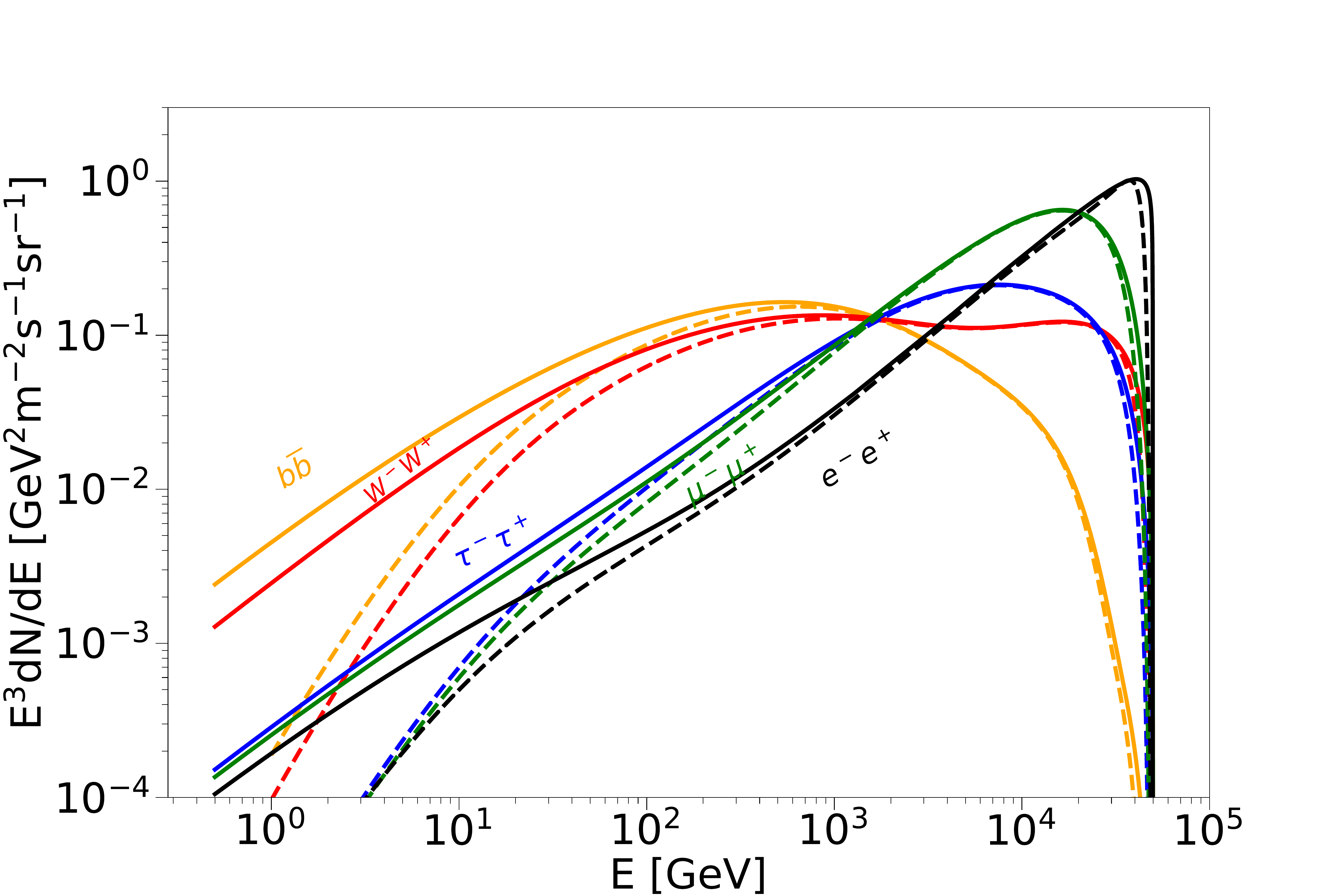}
\caption{The expected electron spectra after propagation originated from DM decay in the Galactic halo for different channels. A DM mass of 100 TeV is assumed here. The dashed lines consider both spatial diffusion and energy loss, while the solid lines neglect the spatial diffusion.}
\label{fig3}
\end{figure}

\subsection{Electrons/positrons from dark matter}
We consider the dark matter particles in the Galactic halo. Dark matter particles may annihilate or decay into SM particles and then produce $\gamma$ photons, neutrinos and electrons/positrons which could be detected by observatories on the Earth. 
For decay dark matter, the electron/positron injection for DM particle mass $m_{\chi}$, energy density $\rho$ and lifetime $\tau$ is given by (in units of ${\rm GeV^{-1}\, cm^{-3} \,s^{-1}}$)
\begin{equation}
\label{dnde}
Q(E,r)=\frac{\rho(r)}{m_{\chi}\tau}\frac{{\rm d}N_{\rm e}}{{\rm d}E_{\rm e}}(E)
\end{equation}
where the ${{\rm d}N_e}/{{\rm d}E_e}$ represents the electron energy spectrum per decay. Since we focus on heavy DM in our work, we use the spectrum calculated by {
\tt HDMSpectra} package \cite{2021JHEP...06..121B}. 

Neglecting the effect of convection and re-acceleration (they are only prominent at low energies), the propagation equation of CR electrons can be expressed as 
\begin{equation}
\frac{{\rm \partial}f}{{\rm\partial}t}=\frac{D(E)}{r^2}\frac{\rm\partial}{{\rm\partial}r}r^2\frac{\rm\partial}{{\rm\partial}r}f+\frac{\rm\partial}{{\rm\partial}E}\left[b(E)f\right]+Q(E,r),
\label{eq:diffeq}
\end{equation}
where $f(E,r,t)$ is the number density of the CR electrons at the position $r$ after propagation. The $Q(E,r)$ and $b(E)$ are the source injection rate and energy loss rate, respectively.

For the energies we are considering ($>$100 TeV), electrons lose energy quickly, so the electrons likely detectable by LHAASO are mainly contributed by dark matter in the vicinity of the Earth. 
{The cooling timescale of a $10\,{\rm TeV}$ ($10\,{\rm PeV}$) electron is $\tau=E/b(E)=31.6\,\rm kyr$ ($0.03\,\rm kyr$). The diffusion length before the electron losing its energy is $\lambda=2\left(\int_{E}^{\infty}{D\left(E^\prime\right)}/{b\left(E^\prime\right)}dE^\prime\right)^{1/2}\simeq833\,{\rm pc}$ ($82\,\rm pc$), which is much smaller than the length scale of the Galactic halo ($r_s=20$ kpc). Within such a small region, the density of DM does not vary significantly (e.g., for $r=7.5$ and $9.5\,{\rm kpc}$, the density are $0.48$ and $0.33\,{\rm GeV/cm^2}$, respectively.).}

We can {therefore} neglect the spatial diffusion $D(E)$ and only consider the energy loss term.
The solution to the propagation equation Eq.~(\ref{eq:diffeq}) can be simplified to (in units of ${\rm GeV^{-1}cm^{-2}s^{-1}sr^{-1}}$)
\begin{equation}
\label{propogation}
\Phi_{e,{\rm DM}}(r,E) =\frac{c}{4\pi}f(E,r)=\frac{c}{4\pi}\frac{1}{b(E)}\int^{m_\chi/2}_{E}dE^{\prime}Q(E^{\prime},r).
\end{equation}
For the energy loss rate, we use the approximation in \cite{Atoian:1995ux}
\begin{equation}
b(E) = b_0+b_1(E/{1\,\rm GeV})+b_2(E/{1\,\rm GeV})^2.
\end{equation}
The $b_0\approx 3\times 10^{-16}\,{\rm GeV\,s^{-1}}$ and $b_1\approx 10^{-15}\,{\rm GeV\,s^{-1}}$ represent the ionization and the bremsstrahlung losses, respectively. The $b_2\approx 1.0\times 10^{-16}\,{\rm GeV\,s^{-1}}$ is the synchrotron and inverse Compton losses for a sum energy density of $1 {\rm\,eV\,cm^{-3}}$ for both the magnetic field and interstellar radiation field. At the energies of $>1\,{\rm GeV}$, the $b_0$ and $b_1$ term can be effectively neglected. As an example, in Fig.~\ref{fig3} we show the spectra after propagation for five different channels. In this figure we set $m_{\chi}$ and $\tau$ to 100 TeV and ${\rm 10^{28}\,s}$.
{To further prove that the diffusion can be effectively neglected, in Fig.~\ref{fig3} we also plot the predicted electron spectra after propagation considering both spatial diffusion and energy loss (see Appendix~\ref{app:b} for the calculation). We can see that compared to the ones ignoring spatial diffusion, they are only different at the low energy end (outside the energy range of interest) which will not affect our results. For simplicity, in the following we will always neglect the spatial diffusion (namely using Eq.~(\ref{propogation})).}

\begin{figure}[t]
\centering
\includegraphics[width=0.45\textwidth]{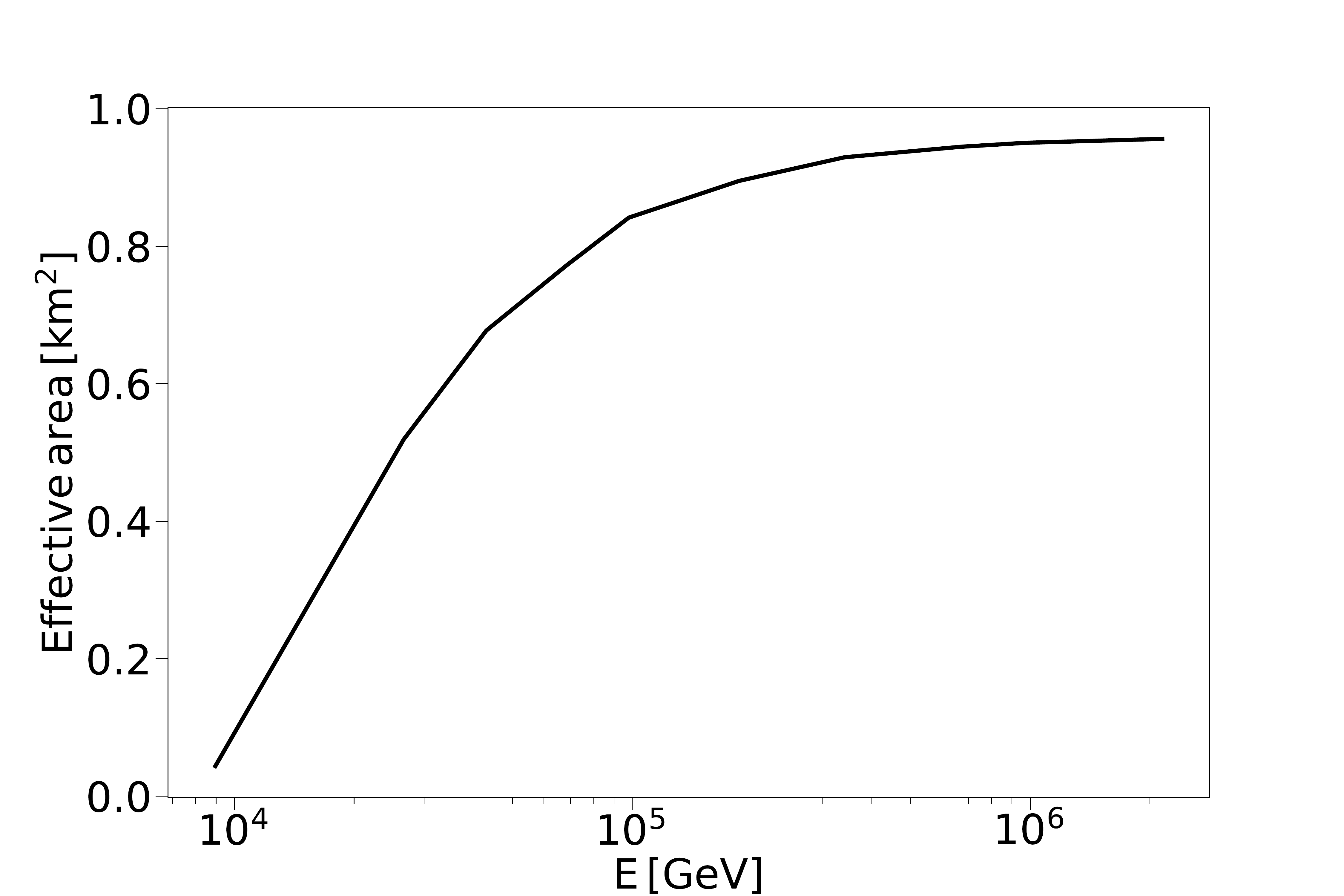}
\caption{Effective area of LHAASO-KM2A as a function of particle energy for gamma-ray/electron observation. This curve is extracted from \cite{bai2022large}.}
\label{fig:aeff}
\end{figure}

\begin{figure*}
\centering
\includegraphics[width=0.45\textwidth]{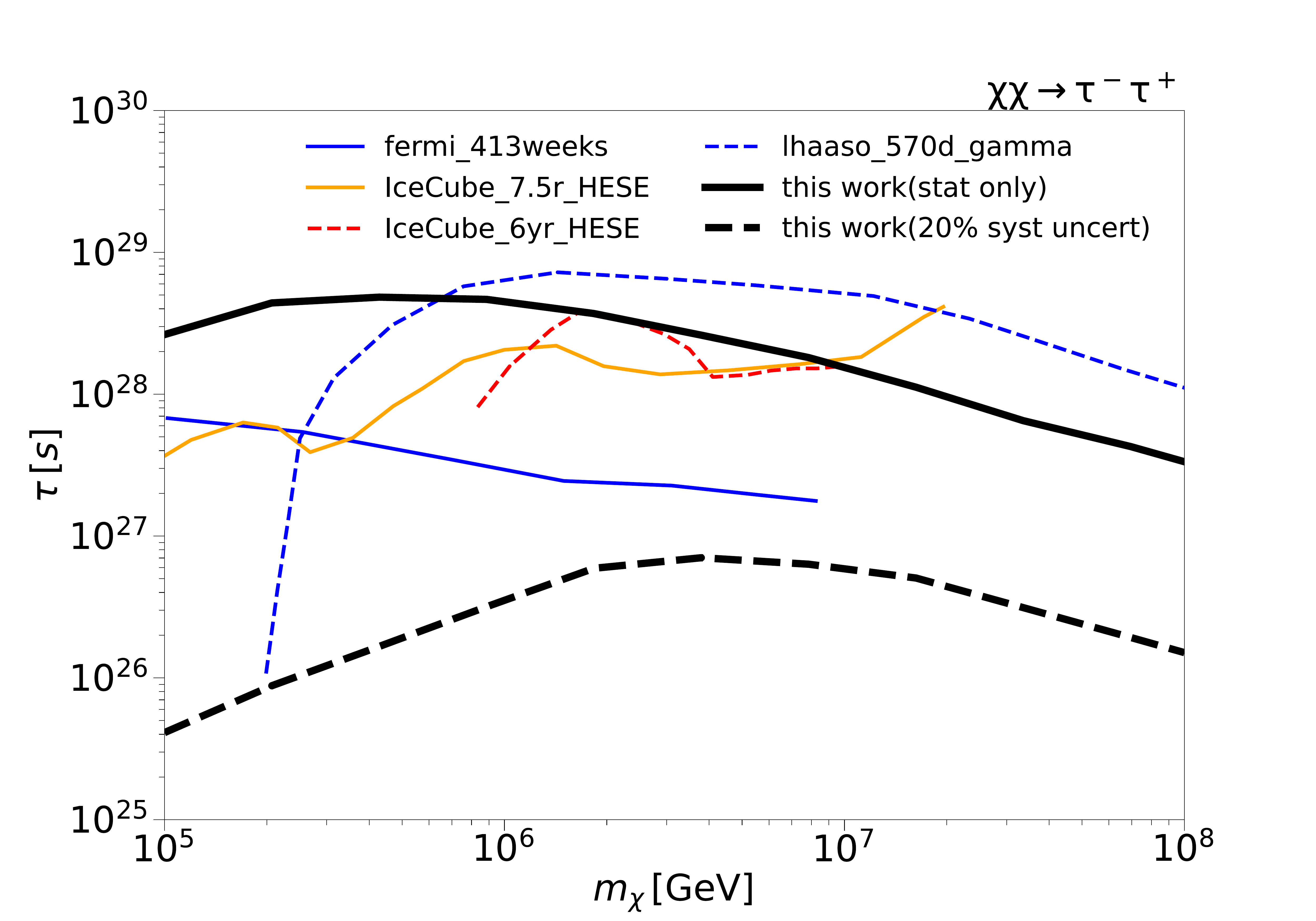}
\includegraphics[width=0.45\textwidth]{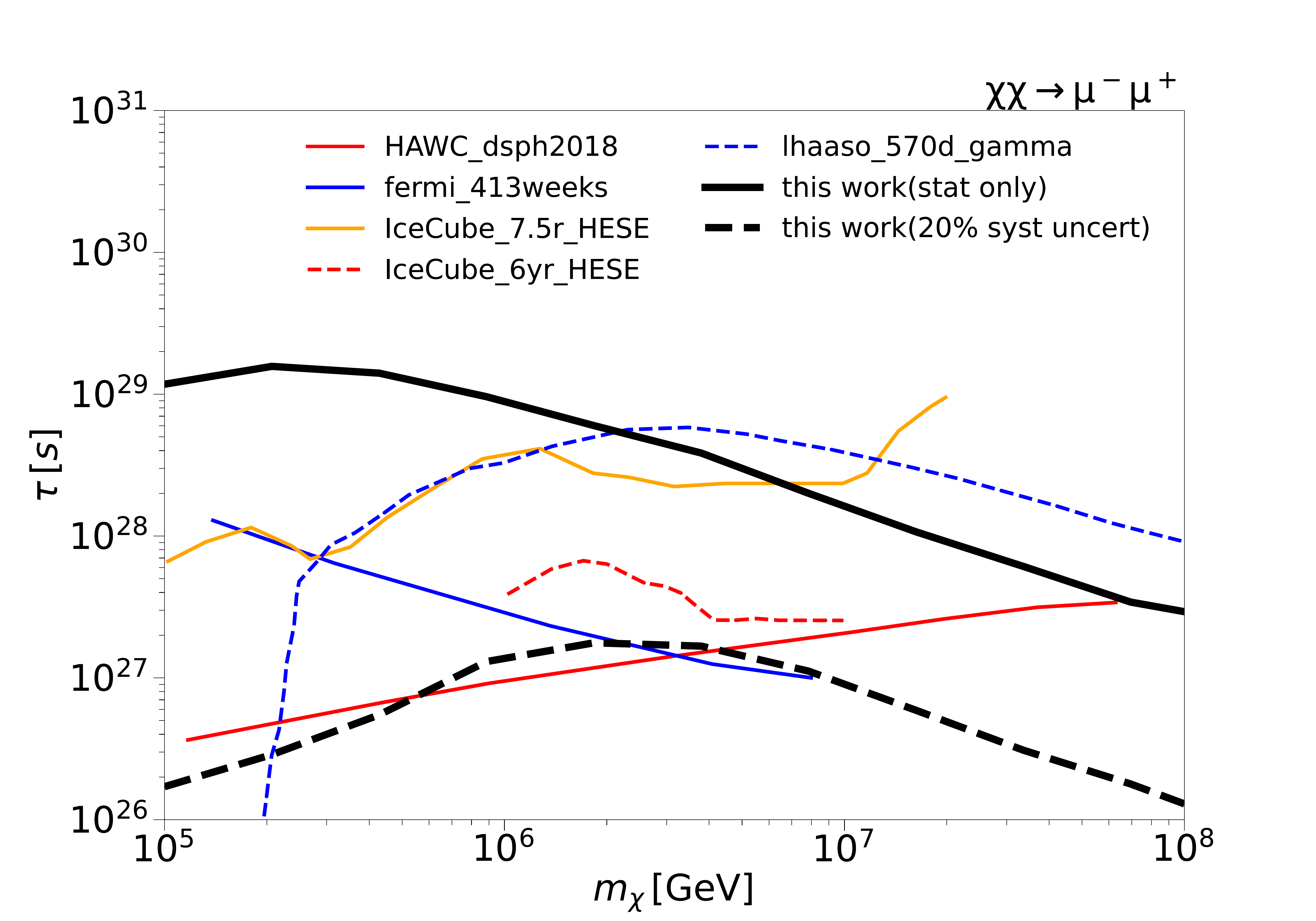}\\
\includegraphics[width=0.45\textwidth]{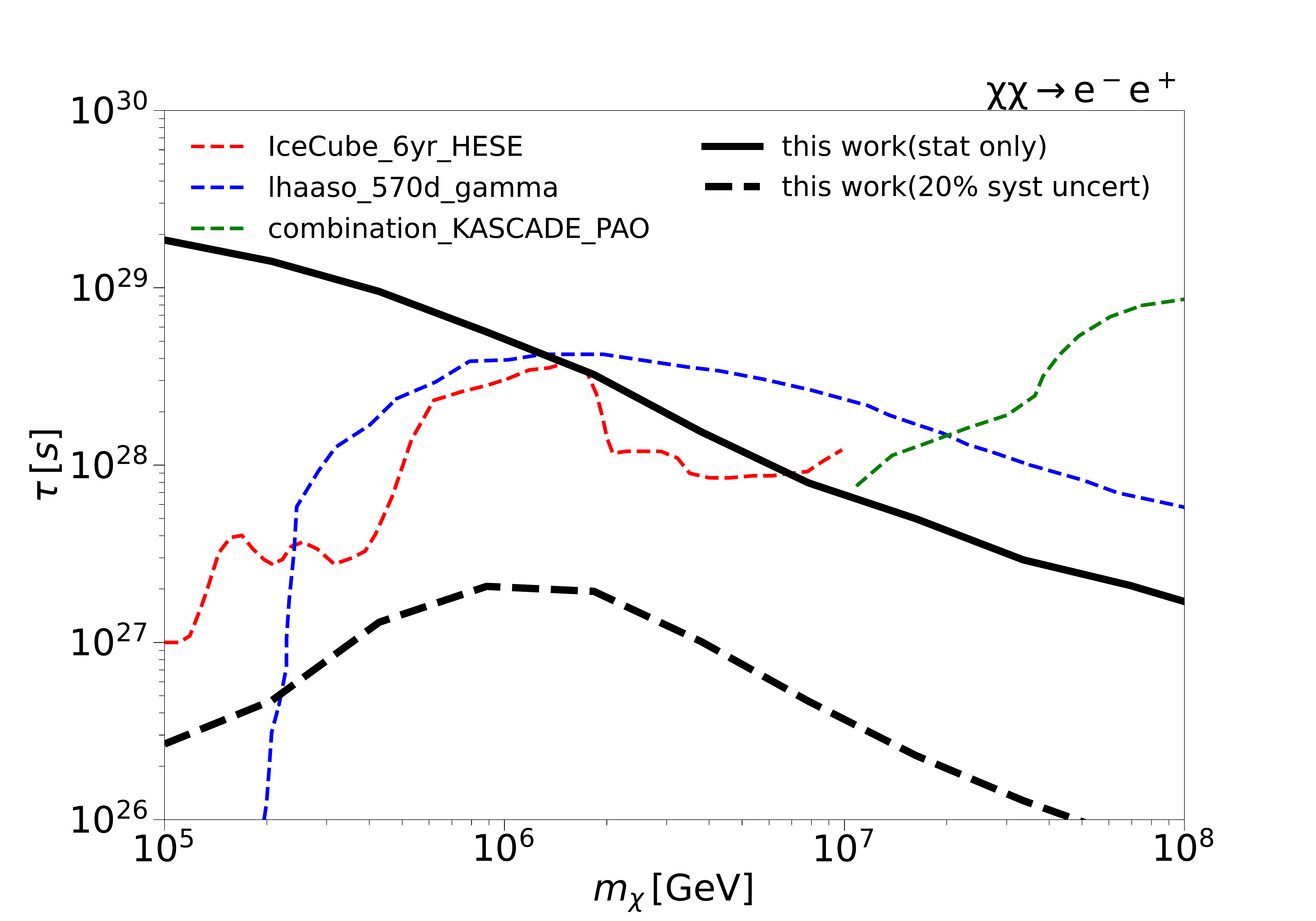}
\includegraphics[width=0.45\textwidth]{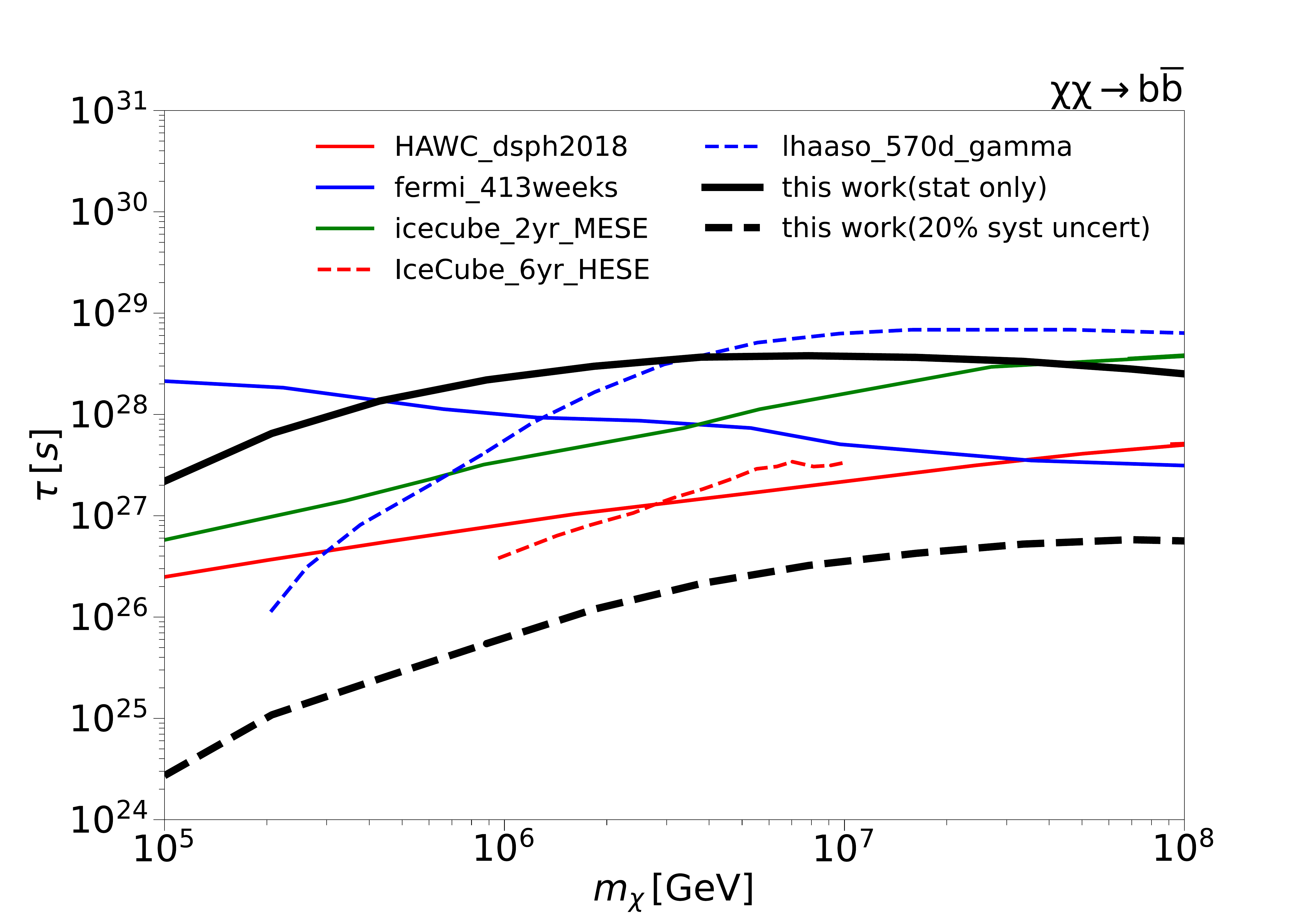}
\caption{The 95\% C.L. lower limits on the lifetime $\tau$ of decaying DM for $\tau^+\tau^-$, $\mu^+\mu^-$, $e^+e^-$ and $b\bar{b}$ channels. 
 The predicted limits that can be placed by the LHAASO electron measurement are shown as black solid and dashed lines, which are the median expected limits derived from the Monte Carlo simulations. Also shown are the limits derived in previous works based on the observations of gamma rays \cite{Cohen:2016uyg,Kalashev:2016cre,Chianese:2021jke,lhaaso22decay} and neutrinos \cite{IceCube:2018tkk,Kachelriess:2018rty,Bhattacharya:2019ucd,IceCube:2022vtr}. }
\label{comparison}
\end{figure*}

Taking into account the contribution from decaying DM, the total expected flux of electrons observed by the LHAASO becomes
\begin{equation}
F_{\rm tot}(E;m_\chi,\tau)=F_{\rm bkg}(E) + \Phi_{e,{\rm DM}}(r_0,E;m_\chi,\tau).
\label{eq:ftot}
\end{equation}
The $r_0=8.5\,{\rm kpc}$ is the position of the Earth, and for the local DM density we use $\rho(r_0)=0.4\,{\rm GeV\,cm^{-3}}$ \cite{Catena:2009mf}. Given that for the energies we are considering ($>10\,{\rm TeV}$) only nearby electrons can reach the Earth before cooling, we don't need to assume the density profile (e.g., NFW \cite{nfw96nfw}, Einasto \cite{einasto}, etc.) of the DM in the Galactic halo. Our results only rely on the local DM density.

\subsection{Monte Carlo simulations and the derivation of lower limits}
To obtain the projected constraints that can be placed by the LHAASO electron measurements, we perform Monte Carlo simulations. We generate pseudo electron spectra observed by the LHAASO {in the energy range 13.3 - 1320 TeV (which are the minimum and maximum energies of the rejection efficiency curve in Fig.~\ref{fig:rej}) based on the background in Sec.~\ref{sec:bkg}. Considering the energy dispersion of LHAASO, we adopt an energy bin size of $\Delta \log(E)=0.2$.}
The expected number of counts in the $k$-th energy bin is given by:
\begin{equation}
\mu_k=\Omega_{\rm fov} T \int_{k} A_{\rm eff} F_{\rm bkg}(E)\,{\rm d}E,
\end{equation}
where the $\Omega_{\rm fov}$ and $A_{\rm eff}$ represent the field of view and effective area of the LHAASO, respectively. We adopt $\Omega_{\rm fov}=2.24\,{\rm sr}$ \cite{bai2022large}. For the effective area, we use the curve in Fig.~\ref{fig:aeff}, which is extracted from \cite{bai2022large}. 
We consider 5 years of measurements, i.e, $T=5\,{\rm yr}$. 
{We caution that this curve is for a normal incidence. The effective area is actually zenith angle dependent and a smaller effective area is expected for larger zenith angle incidence. So our estimate has overestimated the exposure. {As mentioned above, the subtraction of the GP region is also possible to reduce the $\Omega_{\rm fov}$ by 20\%.} However, as will shown below, our final results are systematic uncertainty dominant. The reduce of the statistics by a factor of a few will not change our final conclusion and can be compensated for by extending the observation time.}
Considering Poisson fluctuations, we generate 1000 realizations of the mock spectra (based on the red dashed line in Fig.~\ref{fig2}). Because of LHAASO's large effective area, the Poisson fluctuations can be approximated with Gaussian distribution {(for all the energy bins, the $\mu_k$ is $>10^5$)}. In this work, we sample the observed counts $n_k$ in the $k$-th bin of the mock data according to Gaussian distribution ${\rm \mathcal{N}}(u=\mu_k,\sigma=\sqrt{\mu_k})$ ). 
With these 1000 pseudo spectra on hand, we carry out $\chi_2$ fits to each of the spectra to get best-fit parameters. 
The $\chi^2$ function is defined as
\begin{equation}
\chi^2=\sum\limits_{k}\frac{\left[n_{k}-\mu_{k}(\tau,N_{\rm bkg})\right]^2}{\left(\sqrt{n_{k}} \right)^2},
\end{equation}
{where the DM life time $\tau$ and the normalization parameter $N_{\rm bkg}$ of the background component are free in the fit.}
To find out the best-fit parameters that minimize the $\chi^2$, we use the {\tt iminuit} package \cite{iminuit}.
The $\chi^2$ are computed for both the background-only model (Eq.~(\ref{eq:fbkg})) and the model containing DM component (Eq.~(\ref{eq:ftot})).
For each DM mass $m_\chi$, the 95\% confidence level lower limits on the lifetime $\tau$ correspond to the value leading to $\Delta\chi^2=2.71$.

\section{Results and discussion}
\label{sec:rs}

We scan the DM particle masses $m_{\chi}$ from 100 TeV to about 100 PeV and for each mass we perform the $\chi^2$ fit and derive the lower limits.
The lower limits on $\tau$ for different decay channels are shown in Fig.~\ref{comparison} (thick solid lines). The results for all the channels considered and the 68$\%$ and 95$\%$ containments for the 1000 pseudo observations are demonstrated in Fig.~\ref{result1} (as supplementary results, we also show upper limits on $\left<\sigma v\right>$ for annihilation DM in Fig.~\ref{result2}). As is shown, the LHAASO's electron observation of five years seems to be able to constrain the DM lifetime to the level of $>10^{28}-10^{29}\,{\rm s}$.
For comparisons, also shown in Fig.~\ref{comparison} are previous results on the decaying heavy DM from Refs.~\cite{IceCube:2018tkk,Chianese:2021jke,Cohen:2016uyg,IceCube:2022vtr,Ishiwata:2019aet} based on the observations of neutrinos and $\gamma$ rays. 
As can be seen, for the leptonic channels, the LHAASO's measurement of the CR electrons can improve the current constraints by up to one order of magnitude for $m_\chi<1000\,{\rm TeV}$. 
For the quark channel, the LHAASO's observation can also mildly improve the current results around $m_\chi=1000\,{\rm TeV}$.

However, the above results neglect the systematic uncertainties of the CR electron measurement, which are sizeable for the indirect measurements by ground-based experiments like LHAASO. For gamma-ray observations, the systematic uncertainties have been investigated by studying the variation of the Crab nebula spectrum and the total systematic uncertainty is estimated to be 7\% for the flux \cite{lhaaso2021crabperf}. 
However, the systematic uncertainty of the CR measurement is expected to be much higher than this, around 20\%-50\% (private communication), {which stems from, for example, the poor knowledge of the effective area and the hadron contamination in the electromagnetic showers.}
To account for the systematic uncertainty, we introduce an additional 20\% error into the pseudo spectra (i.e., $\sigma=\sqrt{\mu_k+(20\%\mu_k)^2}$). With the systematic uncertainty included in the calculation, we find that the projected constraints are largely weakened (thick dashed lines in Fig.~\ref{comparison}), weaker than the currently existing constraints for all the DM masses.
{Our analysis suggests that for using the CR electron observation of LHAASO to constrain the DM parameters, effectively reducing the systematic errors would be the key point.}

Although weaker than other experiments (when taking into account the systematic uncertainty), the LHAASO results have an advantage that, in the LHAASO energy range the electrons arriving at the observatory must be produced in the vicinity of the Earth, we do not need to assume the DM density profile of the Galactic halo. The results are therefore less model-dependent.
{Of course, uncertainties in parameters of the local environment still affect our results. 1) This work adopts a DM local density value of $\rho(r_0)=0.4\,{\rm GeV\,cm^{-3}}$ \cite{Catena:2009mf}, there is actually an uncertainty of $\sim0.3-0.7\,{\rm GeV\,cm^{-3}}$ for this value \cite{2010A&A...523A..83S, Iocco:2011jz}. However, recent results point out that most local analyses favor $\rho(r_0)=0.4-0.6\,{\rm GeV\,cm^{-3}}$ \cite{deSalas:2020hbh}, so the density of $0.4\,{\rm GeV\,cm^{-3}}$ used in our work is a relatively conservative value. 2) Uncertainties in the electron energy loss rate would weaken the constraints. The local radiation plus magnetic field density value is possible to be up to $2.6\,{\rm eV\,cm^{-3}}$ \cite{Bergstrom:2013jra}. Adopting this higher value will weaken the results by a factor of few.}

Finally, we need to note that our analysis is based on the information of LHAASO (rejection efficiency, effective area, etc.) publicly reported in the literature, so the results are not guaranteed to be precise enough. In our calculation we adopt the rejection efficiency that is optimized for gamma-ray observation. A dedicated data selection cut for the measurements of electrons is likely to improve the sensitivity.
If the instrument has a rejection power 10 times better than the one currently used, the {\tt 20\% syst} constraints in Fig.~\ref{comparison} can be improved by one order of magnitude.
Regarding the KM2A's detection efficiency (effective area) for hadron cosmic rays, this paper assumes it is the same as the one for electrons/$\gamma$-rays, but in fact there may be some differences between them.
In addition, the analysis of this paper is carried out in a framework that is more commonly used for gamma-ray data, while the data analysis for CRE would be very different.
{We also neglect the energy dispersion, which may bias the results by tens of percent.}
Nevertheless, our results give the first estimation of the DM constraints based on the LHAASO CR measurements and have shown that LHAASO's CRE observations have potential for limiting heavy decaying DM.

\begin{acknowledgments}
We thank Bei Zhou, Huihai He, Yuliang Xin and Shaoqiang Xi for the helpful suggestions and discussions. This work is supported by the National Key Research and Development Program of China (No. 2022YFF0503304) and the Guangxi Science Foundation (grant No. 2019AC20334).
\end{acknowledgments}

\bibliographystyle{apsrev4-1-lyf}
\bibliography{reference.bib}

\widetext
\appendix
\section{Solution to the diffusion equation for point source injection}
\label{app:a}
Considering only the spatial diffusion and energy loss, the diffusion equation of CR electrons can be expressed as 
\begin{equation}
\frac{{\rm \partial}f}{{\rm\partial}t}=\frac{D(E_e)}{r^2}\frac{\rm\partial}{{\rm\partial}r}r^2\frac{\rm\partial}{{\rm\partial}r}f+\frac{\rm\partial}{{\rm\partial}E_e}\left[b(E_e)f\right]+Q(E_e,r,t),
\label{eq:diffeq2}
\end{equation}
For a point source injection, $Q\left(E_e, t\right)\propto E_e^{-\Gamma}\exp(-E_e/E_{\rm cut})$, the above equation can be solved with the Green’s function method,
\begin{equation}
f\left(E_{e}, r, t\right)=\int_0^{t} d t_{0} \frac{b\left(E_{e}^*\right)}{b\left(E_{e}\right)} \frac{1}{\left[\pi \lambda^{2}\left(t_{0}, t, E_e\right)\right]^{{3}/{2}}} \times\exp \left[-\frac{r^2}{\lambda^{2}\left(t_{0}, t, E_e\right)}\right] Q\left(E_{e}^*, t_{0}\right)
\end{equation}
where $r$ is the distance to the source. 
The $E_e^*$ is the initial energy of electron that cool down to $E_e$ in a loss time of $\Delta\tau=t-t_0$:
\begin{equation}
E_e^*= \frac{E_{e}}{\left[1-b_{2} E_{e}\left(t-t_{0}\right)\right]},
\end{equation}
and the diffusion length is given by (assuming $b(E_e)\simeq b_2E_e^2$)
\begin{equation}
\lambda(E)= 2\sqrt{\frac{D_0[E_e^{*(\delta-1)}-E_e^{(\delta-1)}]}{ b_2(\delta-1)}}.
\end{equation}

\section{Solution to the diffusion equation for NFW injection}
\label{app:b}
For the injection with an NFW distribution, the solution is
\begin{align}
N\left(E_e, r, t\right)=&\int_0^t d t_0 \frac{b\left(E_e^*\right)}{b\left(E_e\right)} \frac{1}{\left[\pi \lambda^2\left(t_0, t, E_e\right)\right]^{{3}/{2}}} \int \exp \left[-\frac{\left|\vec{r}-\vec{r}^{\prime}\right|^2}{\lambda^2\left(t_0, t, E_e\right)}\right] Q\left(E_e^*, \vec{r}^{\prime}, t_0\right) \mathrm{d} \vec{r}^{\prime} \nonumber\\
=& \int_0^t d t_0 \frac{b\left(E_e^*\right)}{b\left(E_e\right)} \frac{1}{\left[\pi \lambda^2\left(t_0, t, E_e\right)\right]^{3 / 2}} \int  \frac{\pi \lambda^2 r^{\prime}}{r}\left\{\exp \left[-\frac{\left(r-r^{\prime}\right)^2}{\lambda^2}\right]-\exp \left[-\frac{\left(r+r^{\prime}\right)^2}{\lambda^2}\right]\right\}Q\left(E_e^*, r^{\prime}, t_0\right) \mathrm{d} r^{\prime}.
\end{align}
The injection rate is
\begin{equation}
\label{dnde}
Q(E,r,t)=\frac{\rho(r)}{m_{\chi}\tau}\frac{{\rm d}N_{\rm e}}{{\rm d}E_{\rm e}}(E),
\end{equation}
where the ${{\rm d}N_e}/{{\rm d}E_e}$ represents the electron energy spectrum per decay.

\newpage
\section{Results for various channels}
\begin{figure*}[h]
\centering
\includegraphics[width=0.45\textwidth]{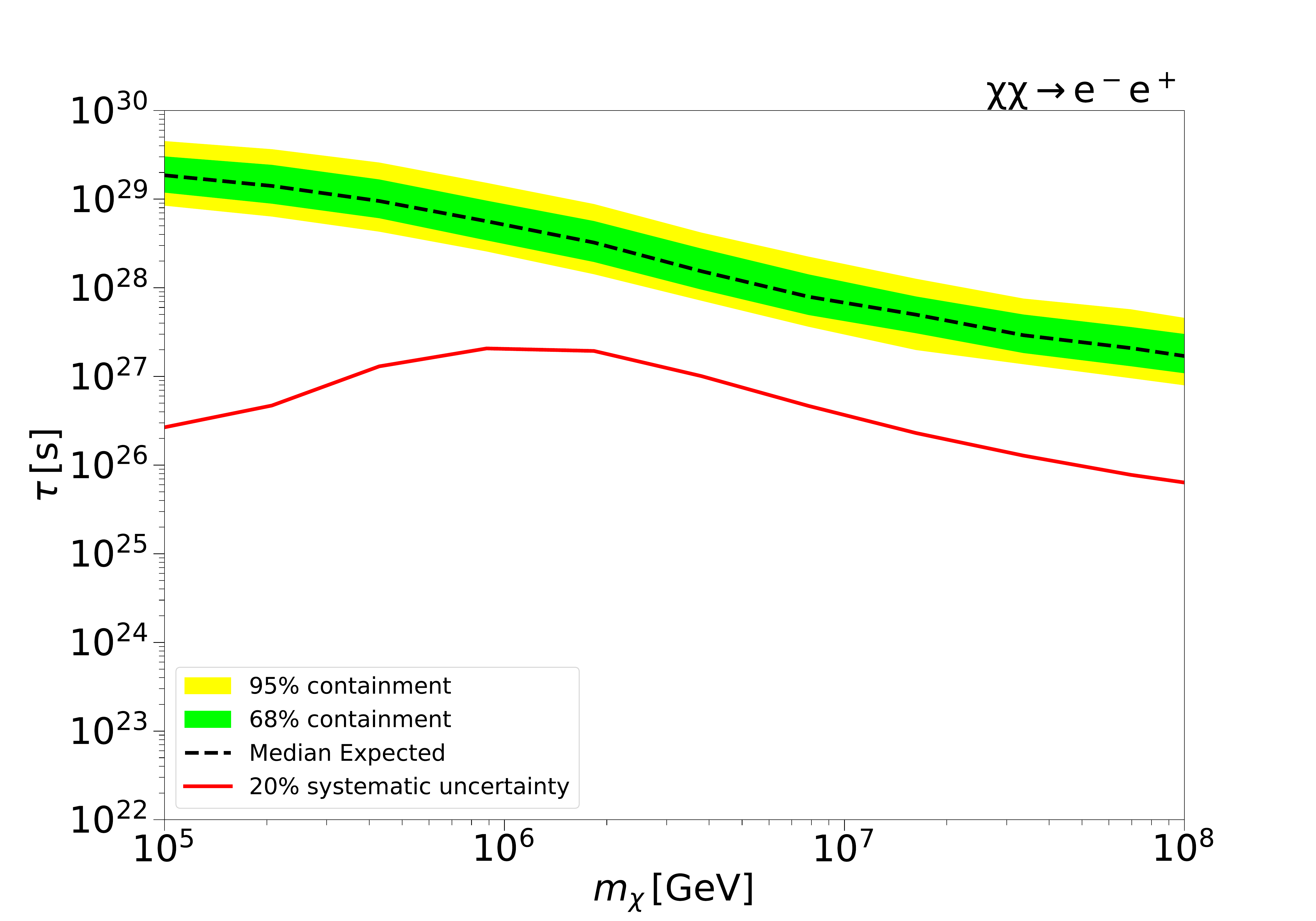}
\includegraphics[width=0.45\textwidth]{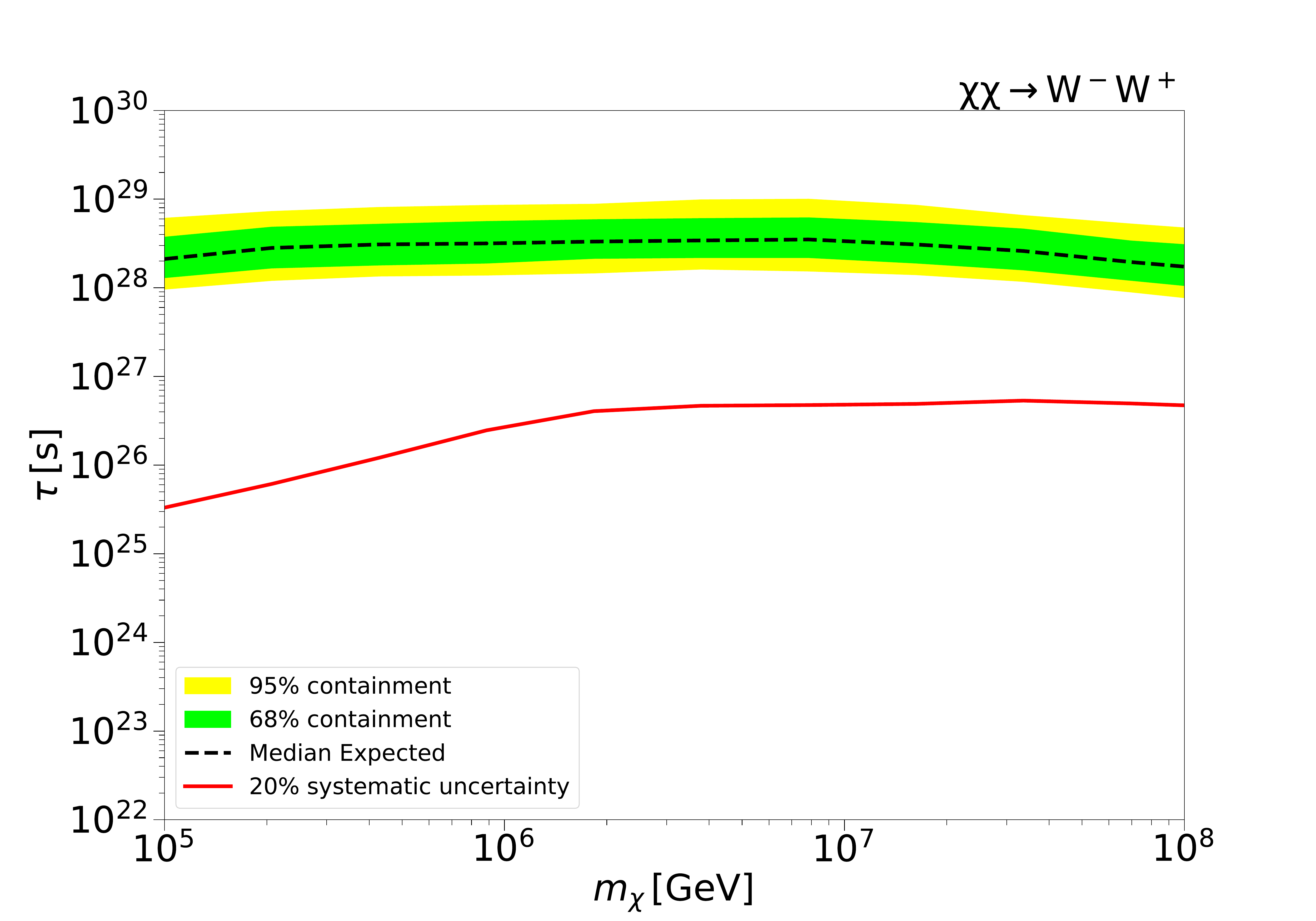}\\
\includegraphics[width=0.45\textwidth]{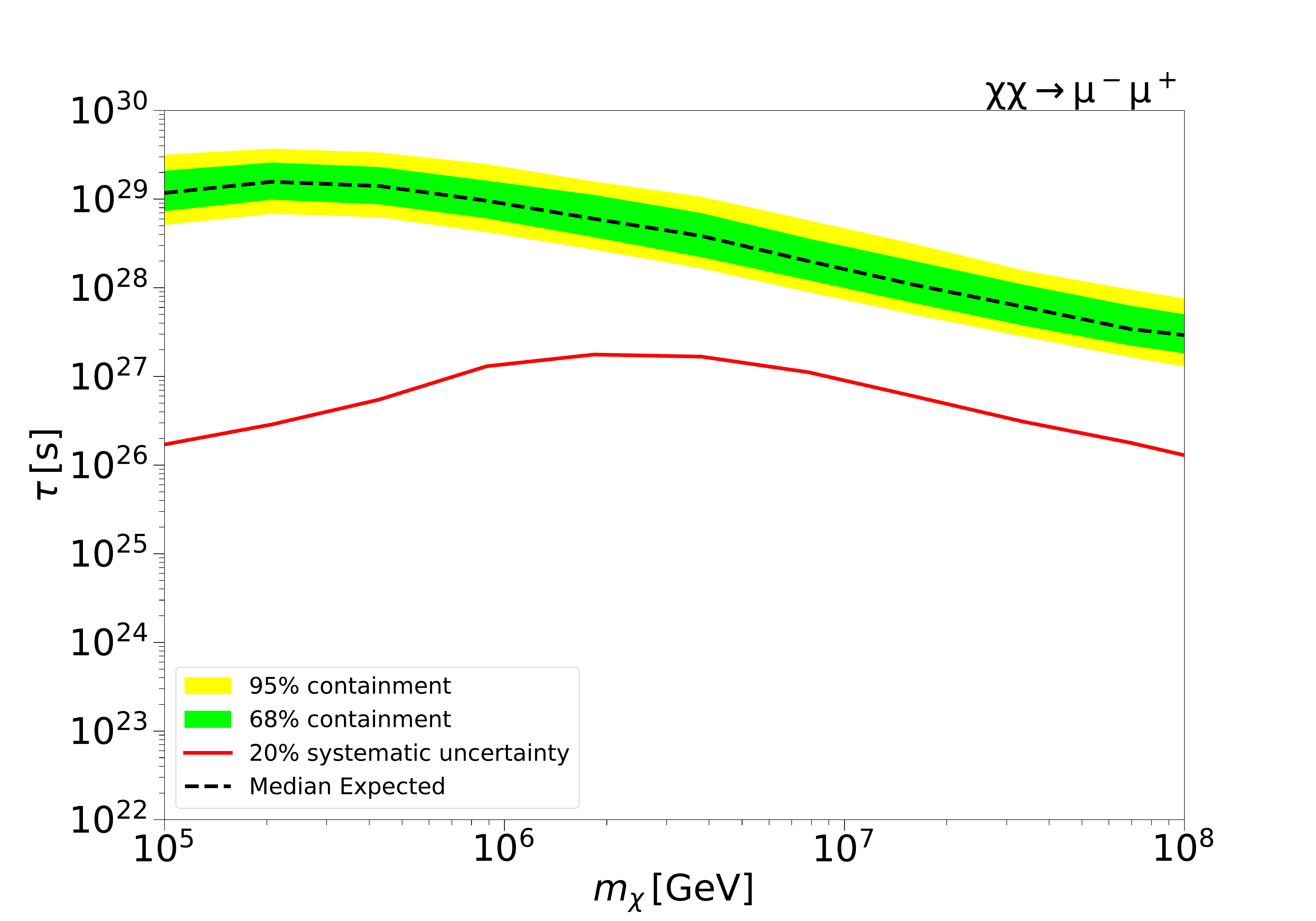}
\includegraphics[width=0.45\textwidth]{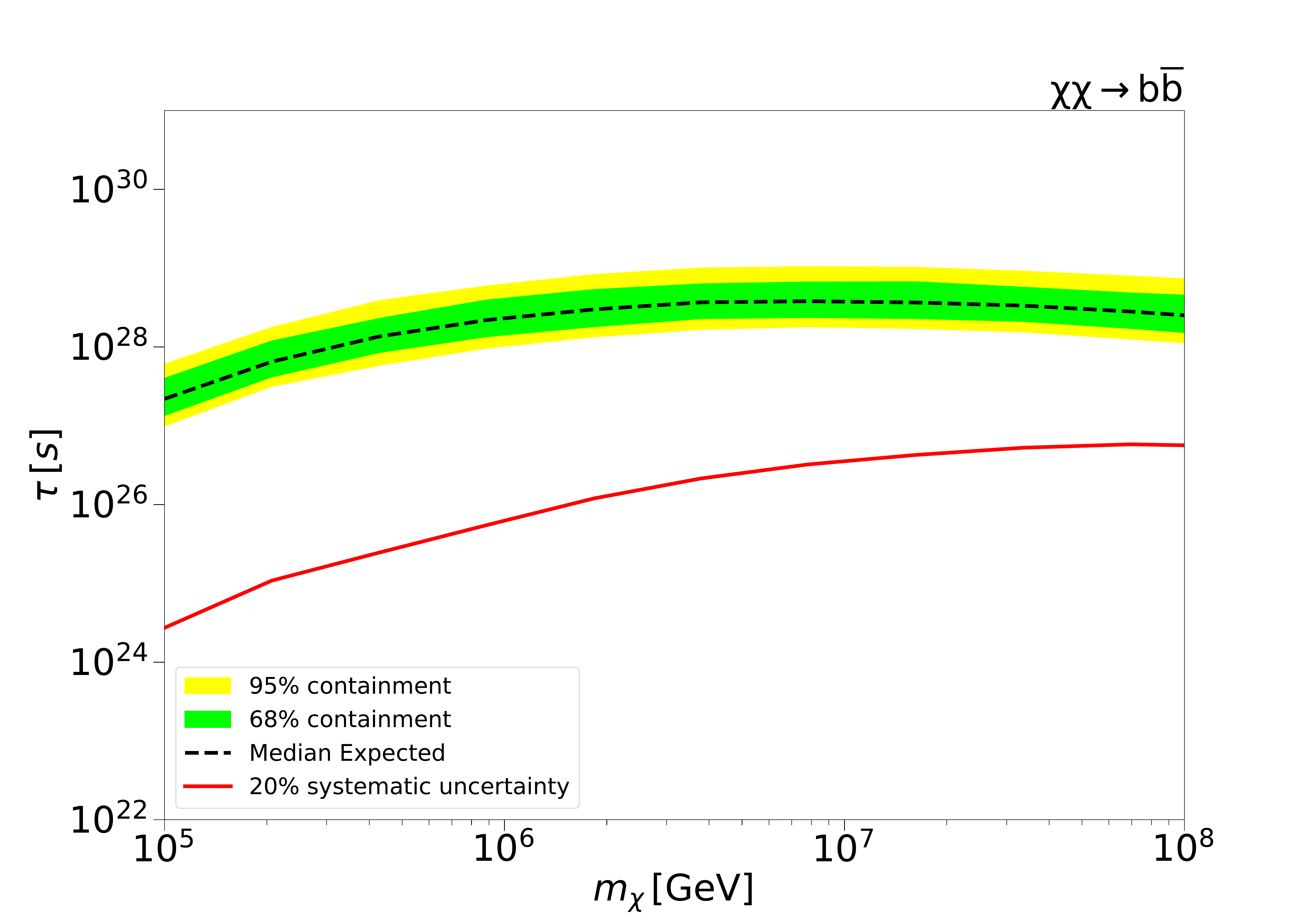}\\
\includegraphics[width=0.45\textwidth]{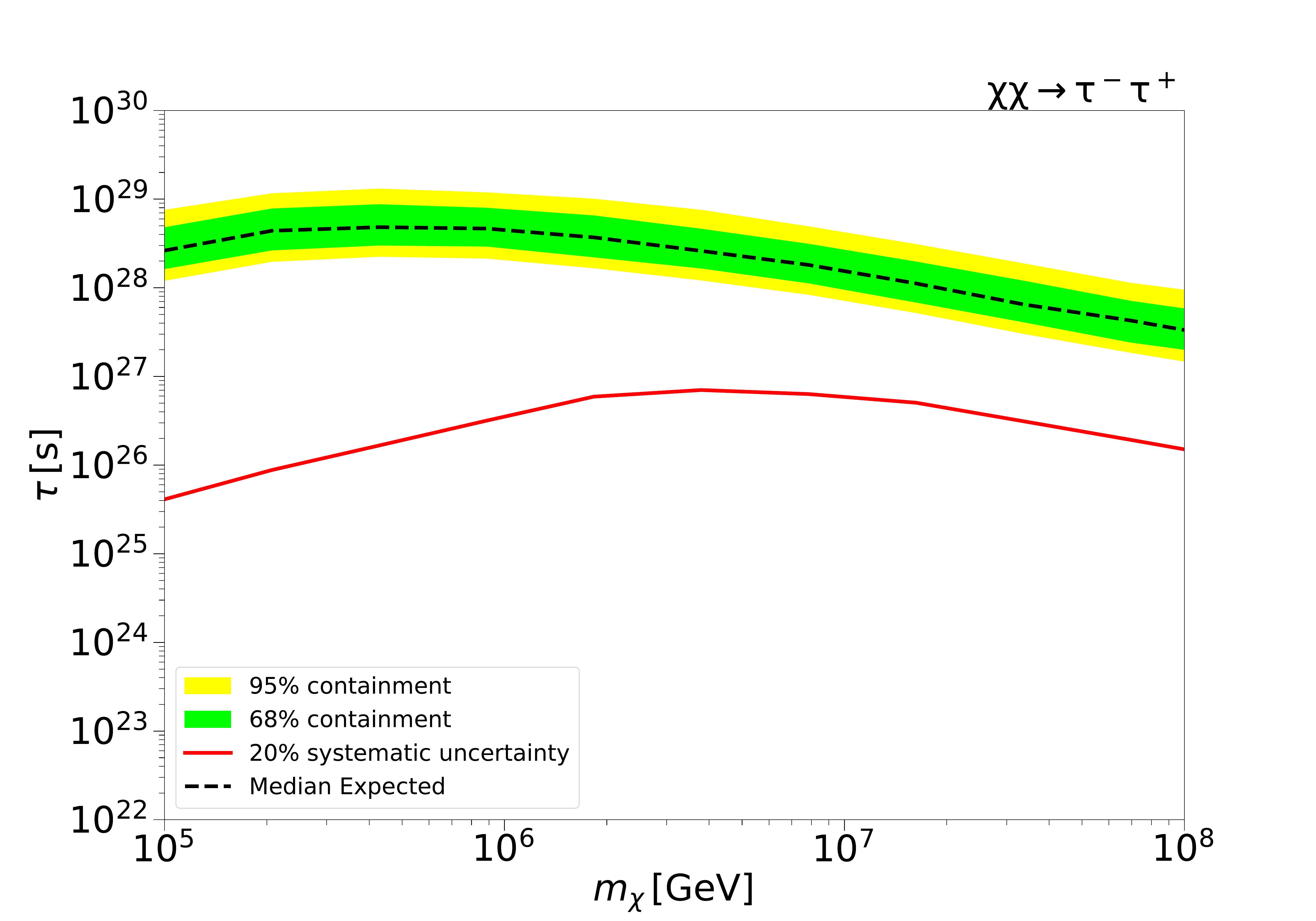}
\includegraphics[width=0.45\textwidth]{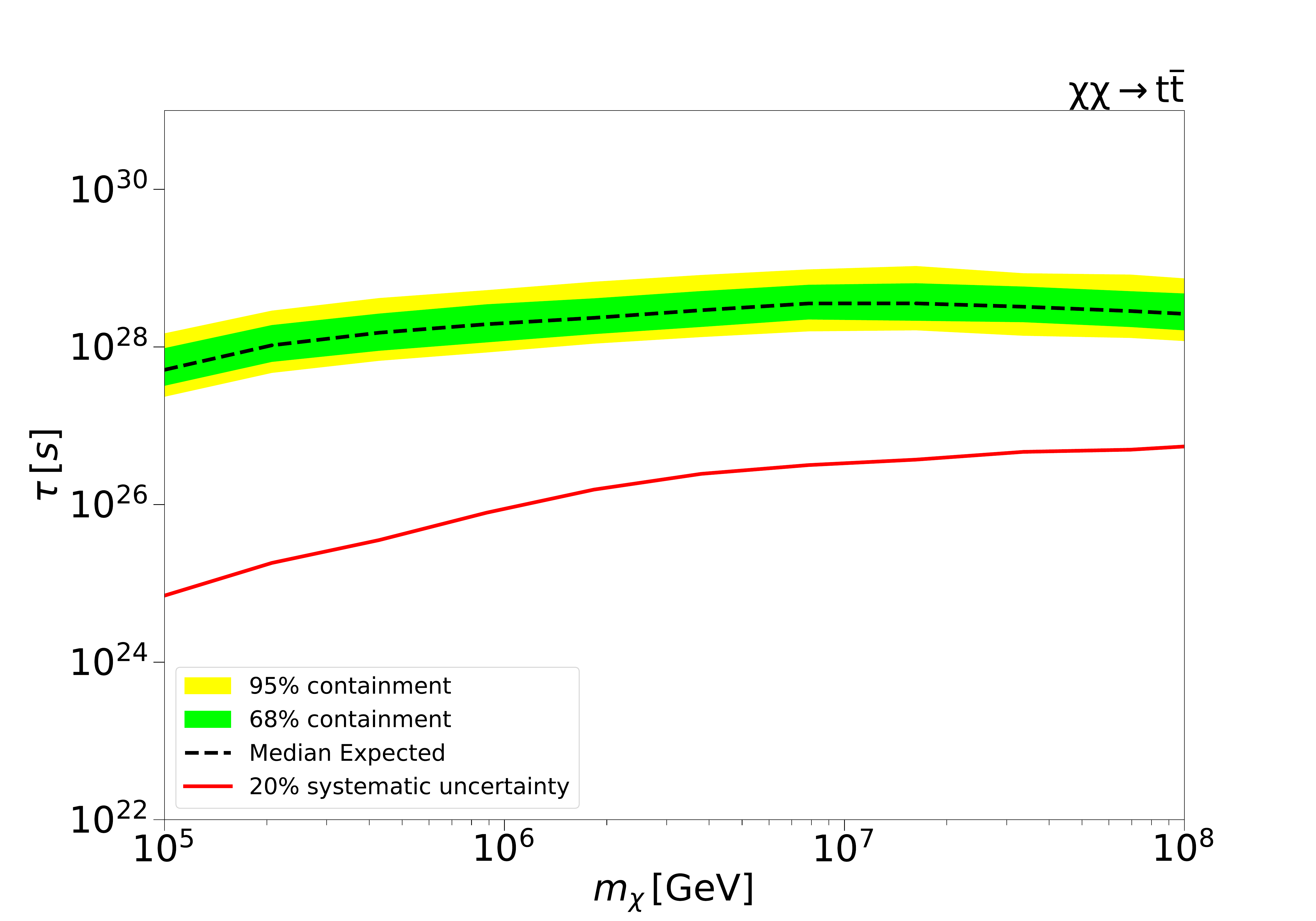}
\caption{The 95\% C.L. lower limits on lifetime $\tau$ of decaying DM for various channels. Yellow (green) bands show the 68\% (95\%) expected containments derived from 1000 Monte Carlo simulations.}
\label{result1}
\end{figure*}

\begin{figure*}[h]
\centering
\includegraphics[width=0.45\textwidth]{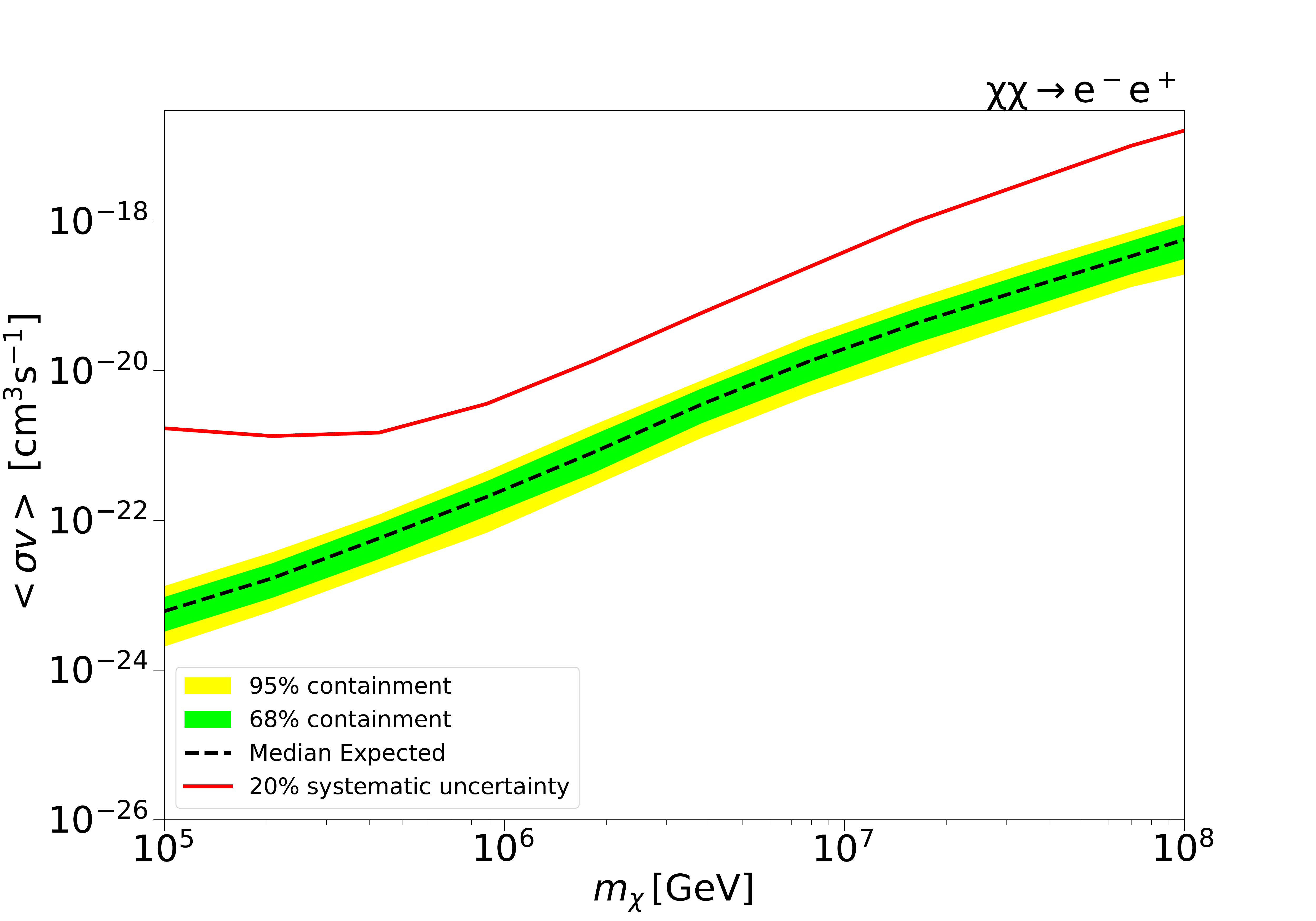}
\includegraphics[width=0.45\textwidth]{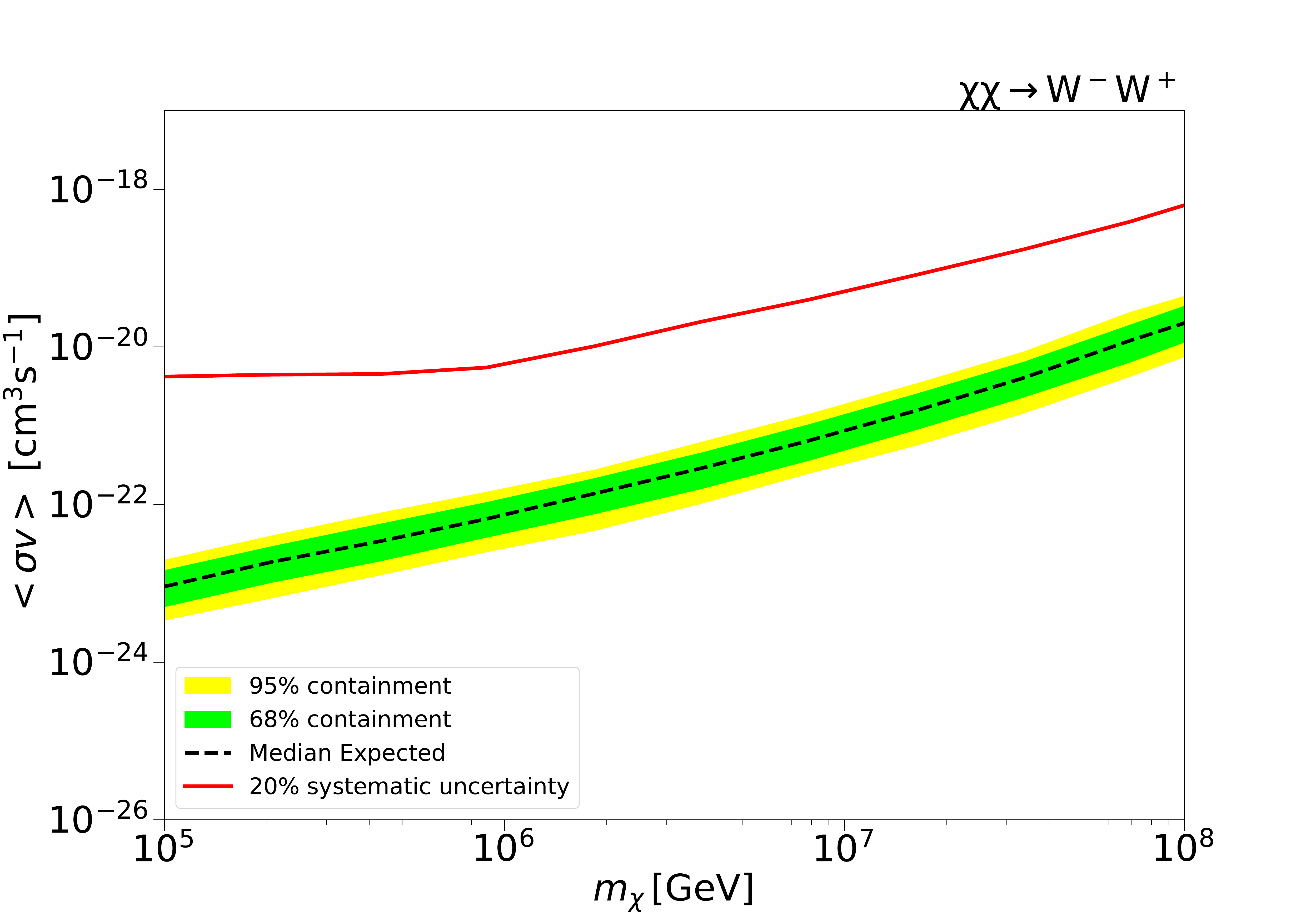}\\
\includegraphics[width=0.45\textwidth]{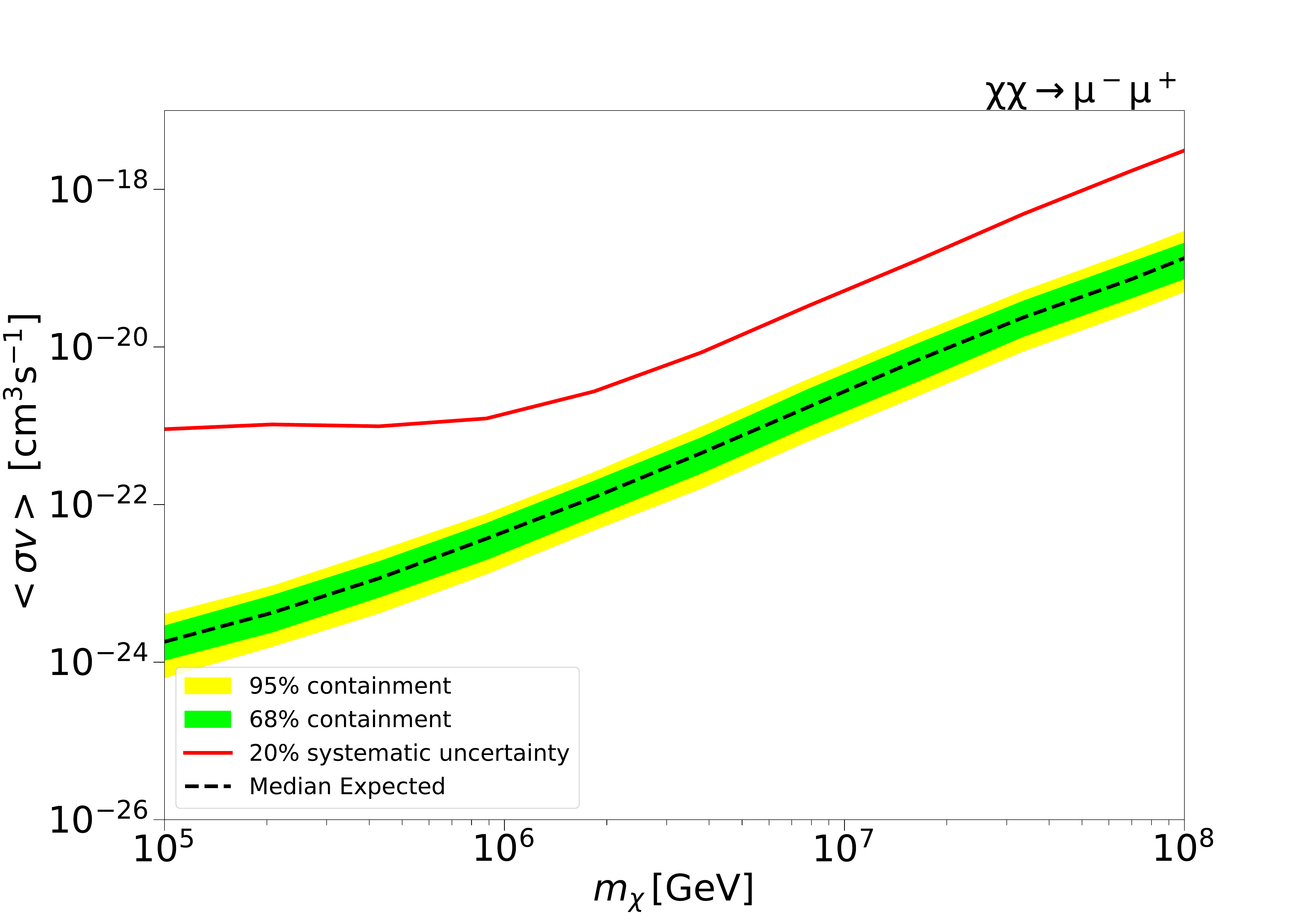}
\includegraphics[width=0.45\textwidth]{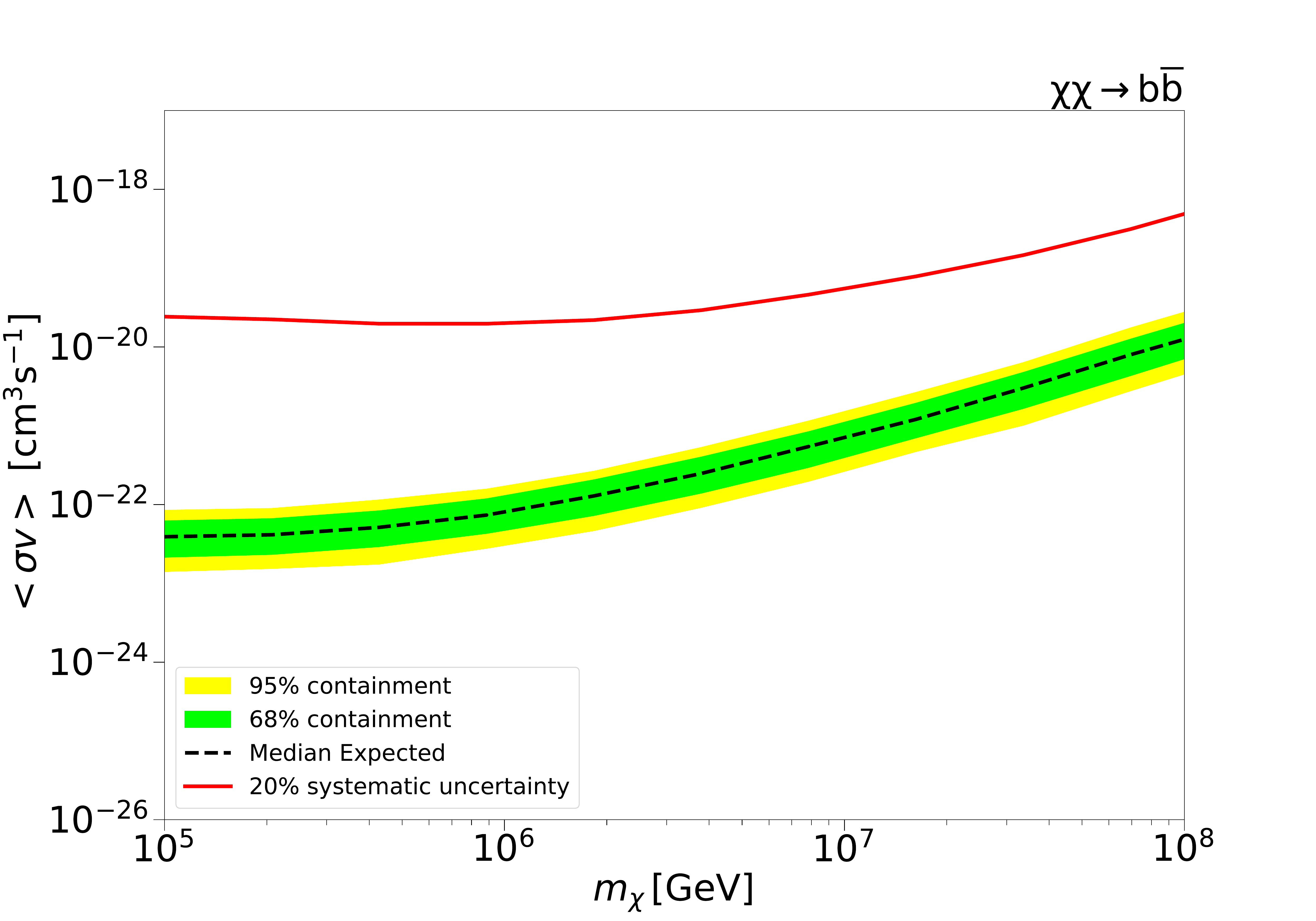}\\
\includegraphics[width=0.45\textwidth]{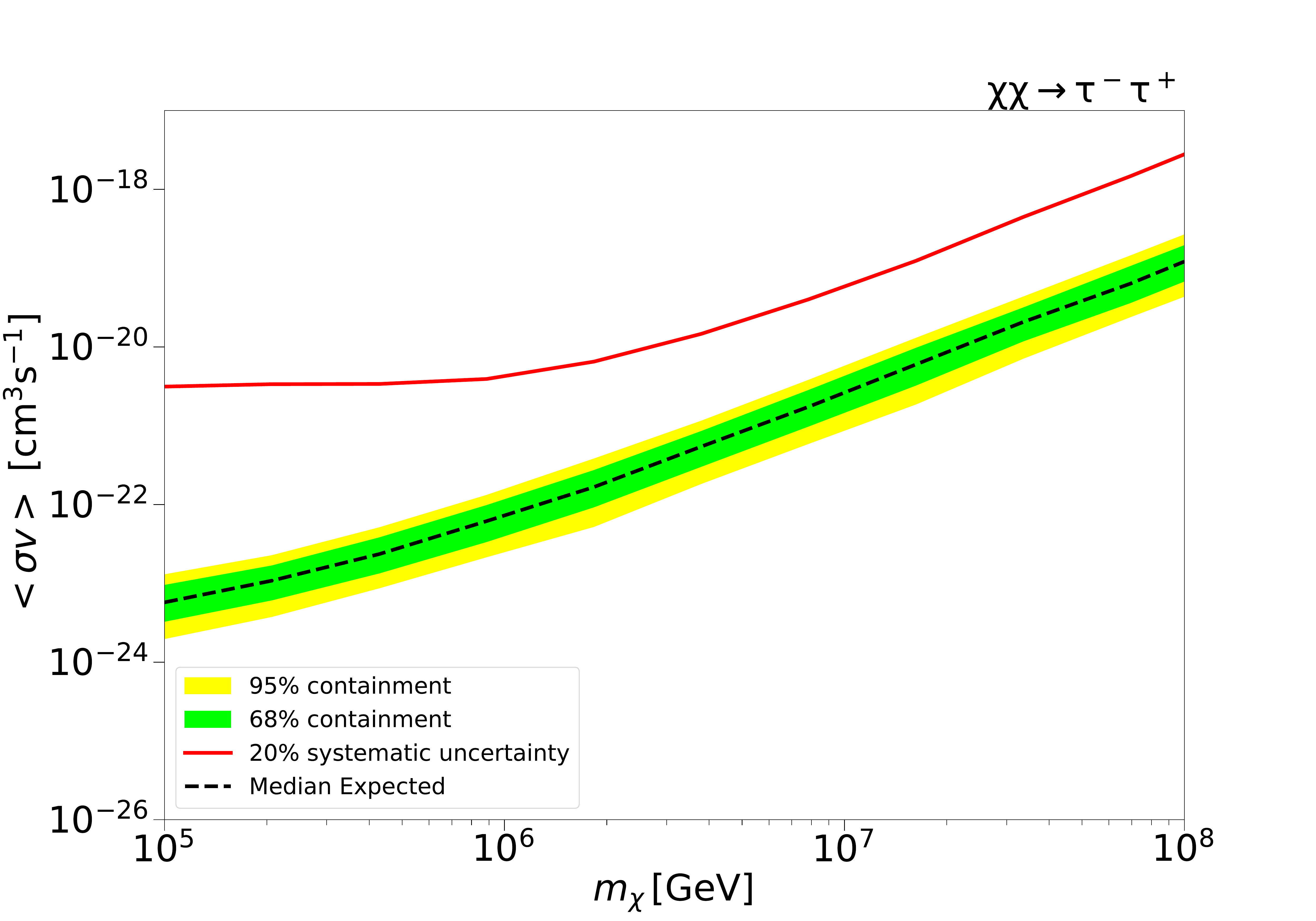}
\includegraphics[width=0.45\textwidth]{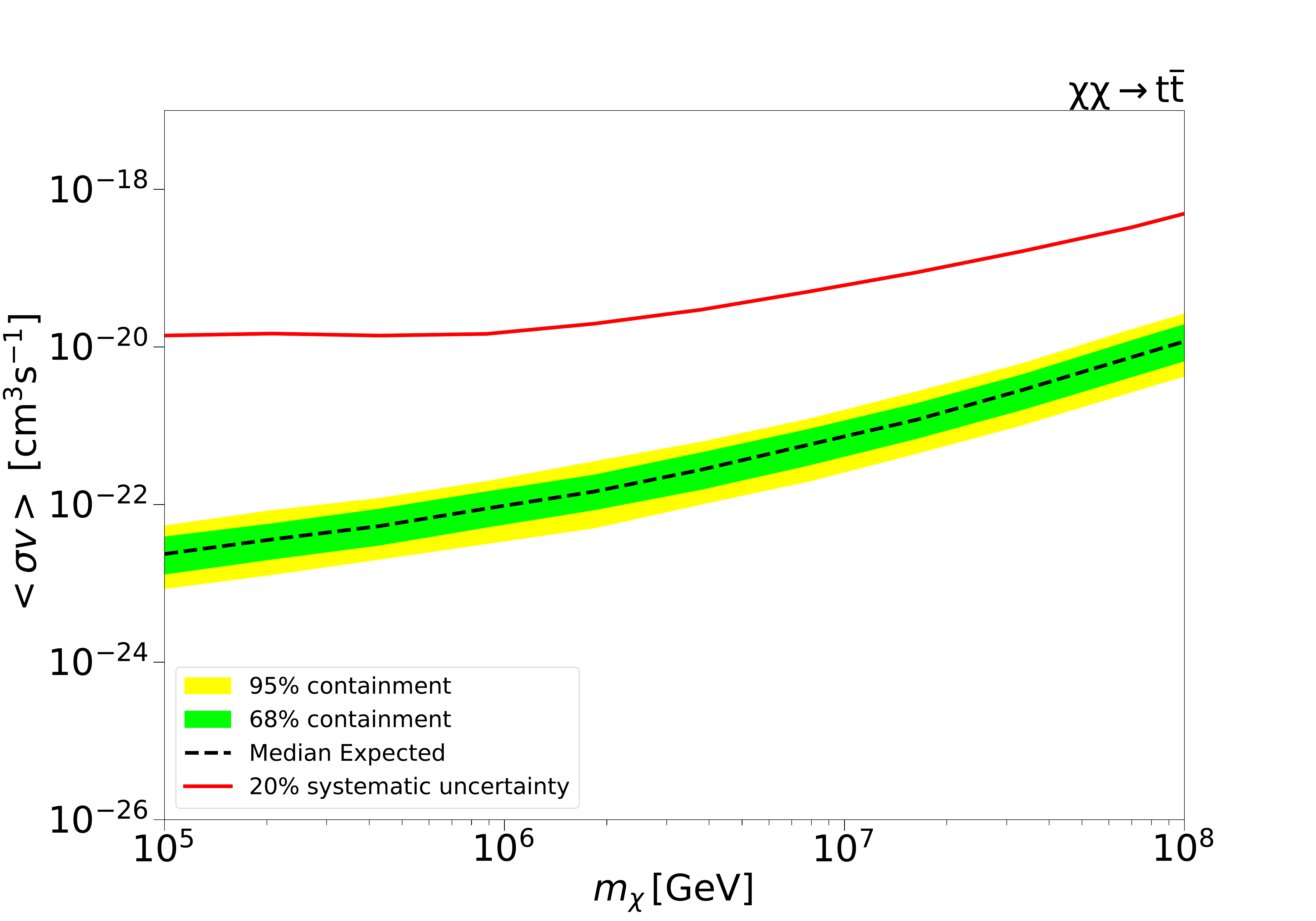}
\caption{The 95\% C.L. upper limits on cross section $\left<\sigma v\right>$ of annihilation DM for various channels.}
\label{result2}
\end{figure*}

\end{document}